\documentclass{aa_edited}

\usepackage{float}
\usepackage{graphicx}
\graphicspath{{figures/}}
\usepackage{txfonts}
\usepackage{amsmath, bm}
\usepackage{amssymb}
\usepackage{mathtools}
\usepackage{nccmath}
\usepackage{booktabs}
\usepackage{microtype}
\usepackage{subcaption}
\usepackage[dvipsnames]{xcolor}
\definecolor{petrol}{RGB}{20, 150, 160}
\definecolor{midori}{HTML}{2A603B}
\usepackage{natbib}
\bibpunct{(}{)}{;}{a}{}{,}
\usepackage{hyperref}
\definecolor{linkcolor}{rgb}{0.0,0.3,0.5}
\hypersetup{
 colorlinks=true,
 citecolor=blue,
 linkcolor=blue,
 urlcolor=blue,
 filecolor=blue,
 pdftitle={Modeling the nHz gravitational wave background with t-process},
 }
\usepackage{adjustbox}
\usepackage{multirow}
\usepackage{colortbl}
\usepackage[normalem]{ulem}

\begin{document}
\title{Modeling the nanohertz gravitational wave background with the t-process: Improved frequentist analysis of gravitational-wave power}
\titlerunning{Modeling the nHz GWB with the t-process}
\authorrunning{Moreschi et al.}

\author{B. E. Moreschi \inst{1,2,3}\fnmsep\thanks{e-mail: \texttt{b.moreschi1@campus.unimib.it}} \and
K. A. Gersbach \inst{4} \and
G. Shaifullah \inst{1,2,3} \and
S. R. Taylor \inst{4} \and
A. Sesana \inst{1,2,4}}

\institute{Dipartimento di Fisica “G. Occhialini”, Università degli Studi di Milano-Bicocca, Piazza della Scienza 3, I-20126 Milano, Italy \and
INFN, Sezione di Milano-Bicocca, Piazza della Scienza 3, I-20126 Milano, Italy \and
INAF – Osservatorio Astronomico di Cagliari, Via della Scienza 5, I-09047 Selargius (CA), Italy \and
INAF – Osservatorio Astronomico di Brera, Via Brera 20, I-20121 Milano, Italy \and
Department of Physics \& Astronomy, Vanderbilt University, 2301 Vanderbilt Place, Nashville, 37235, TN, USA\\
}

\date{Received xxxx; accepted xxxx}

\abstract{Pulsar timing array (PTA) collaborations have recently reported evidence for a gravitational wave background (GWB) signal in the nanohertz band, which most likely originates from the incoherent superposition of GWs emitted by a population of supermassive black hole binaries (SMBHBs). Since the number of binaries emitting in each frequency bin is finite, and since individual loud sources can dominate the GWB, the power spectral density (PSD) of the background is expected to deviate significantly from the smooth power law that describes an idealized population. PTA searches usually describe the PSD either with a power law or with a free spectrum model, which fits the PSD in each frequency bin as a free parameter. In this paper, we explore the t-process as a third model; it assumes a power law as the underlying spectrum but leaves each frequency bin free to deviate from it. We compare the performance of these three models on simulated datasets, injecting $100$ realizations of a realistic SMBHB population, to evaluate how well each model recovers a complex PSD. Using Bayesian and frequentist tools, we find that the t-process model gives the most precise PSD results, with uncertainties $43\%$ and $26\%$ narrower than those obtained with the power law and the free spectrum, respectively, while the improvement in accuracy is smaller. Moreover, using the t-process model, we compute isotropic and anisotropic signal-to-noise ratios with median gains of about $46\%$ over the power law and $65\%$ over the free spectrum in the first nine frequency bins. The t-process also returns lower anisotropy p-values and among the frequency bins with $p<0.01$, it allows to localize the loudest binary within the angular resolution of the sky maps in $14.3\%$ of the cases, compared to the $10\%$ for the power law and the $6.2\%$ for the free spectrum.}


\maketitle

\section{Introduction}
\label{sec:intro}

In recent years, {evidence at different significance levels for a gravitational-wave background} (GWB) in the nanohertz (nHz) regime has been reported by multiple pulsar timing array (PTA) collaborations, including the North American Nanohertz Observatory for Gravitational Waves \citep[NANOGrav;][]{2023ApJ...951L...8A}, the European and Indian collaborations \citep[EPTA and InPTA;][]{EPTA_DR2}, the Parkes PTA \citep[PPTA;][]{PPTA_gwb}, the Chinese PTA \citep[CPTA;][]{CPTA_gwb} and the MeerKAT PTA \citep[MPTA;][]{meerkat}. Moreover, a comparison of different datasets shows that the results are {mutually consistent} \citep{IPTA2023_Baker}.

In PTA experiments, the stochastic GWB, expected to be the dominant signal in the nHz regime, {appears in the times of arrival (ToAs)} of rapidly spinning millisecond pulsars as low-frequency red noise that is common to all pulsars of the array. The signal itself is correlated between pulsars {as a function of their angular separation on the sky}, following the Hellings and Downs (HD) function, as predicted by \citet{HD}.

Although the most likely origin of this signal is the incoherent superposition of the GWs emitted by a population of supermassive black hole binaries (SMBHBs) forming after galaxy mergers \citep{Rajagopal_1995, jb2003, wl2003, high_freq, rosado, bence_exp}, the observed GWB can also be produced by cosmic strings \citep{1976JPhA....9.1387K, 2000PhRvL..85.3761D}, inflationary scenarios \citep{bartolo, boyle, sorbo}, cosmic domain walls \citep{hiramatsu}, primordial curvature perturbations \citep{tomita, matarrese}, or by first-order phase transitions in the early universe \citep{kosowsky, hindmarsh}. {Determining which of these mechanisms produces the observed signal is therefore a major focus of the field.}

Among the properties that can help {identify the origin of the signal} are its spectral shape and its power distribution in the sky.
Considering an idealized population of circular SMBHBs whose evolution is driven by GW emission only, the characteristic strain is expected to follow a power law (PL) as described in \citet{phinney}. However, for a realistic population, the number of binaries emitting in each frequency bin is finite and, especially at high frequencies, the spectrum {can deviate significantly} from this PL approximation. Exceptionally loud binaries can add power in the specific bins in which they emit and even become individually resolvable \citep{sesanavecchiovolonteri, ravi2012, kelley2017, B_2023, Ferranti_2025, b_2025}.
In addition, the coupling between the binaries and their stellar or gaseous environment, as well as the eccentricity of their orbits, tend to suppress power at low frequencies and {flatten the spectrum} \citep{Sesana_2013, ravi, sampsoncornish, 2024MNRAS.532..295F, Valtolina_2024}. Detecting these deviations from a PL would be a strong hint in favor of the SMBHB interpretation.

{The discrete nature of the source population can also make the GWB anisotropic.} A background produced by a finite number of SMBHBs is not uniformly distributed in the sky, and the loudest binaries produce hot spots on the GW power maps \citep{Mingarelli_2013, taylor, rut}. Several methods borrowed from the analysis of the cosmic microwave background (CMB) have been adapted to PTA data to describe these features \citep{Olmez_2012, Gair_2014, Mingarelli_2017}, and searches for anisotropy have been carried out on real datasets with so far inconclusive results \citep{Taylor_2013, NG15aniso, Grunthal_2024}. The sensitivity of current and future arrays to anisotropy has also been studied in detail \citep{ptr22, depta2024}, since detecting it and resolving an individual binary remain among the main goals of the field \citep{ng15yrCGW, EPTA_5, meerkat}. 

Inferring this information from the data requires a model that adequately describes the GWB power spectrum. The two standard models used in the literature are the PL and the free spectrum (FS). The former cannot detect any deviation from its own shape, since it is described by its amplitude and spectral index, and it has an inherent relation of power between different frequency bins. The latter fits the power in each frequency bin as a free parameter and therefore can model deviations from the PL. However, it analyzes each bin in isolation, ignoring by construction any possible correlation with other bins. The {putative SMBHB origin} of the signal suggests an alternative description, falling in {between} these two. Since different binaries emit at different frequencies, the power in each bin should be free to {deviate} from a continuous analytic function. At the same time, the fact that each individual SMBHB belongs to a collective cosmic population suggests some underlying regularity in the overall functional form of the signal. 
A model that can better accommodate the nature of a GWB sourced by a cosmic population of SMBHBs is the t-process (TP). This model uses a PL as the underlying description of the spectrum but introduces a per-bin parametric deviation from it, offering a compromise between continuity and flexibility. The TP was originally introduced to describe an excess of power in the red noise of a single pulsar \citep{2019ApJ...880..116A, 2020ApJ...900..102A} and was later applied to the common GWB signal \citep{2023ApJ...951L...8A}. \citet{2024ApJ...976..212S} studied the TP systematically, comparing it with the PL and the FS on simulated datasets, and showed that the TP recovers the injected power spectral density (PSD) even when it deviates from a PL. The TP is therefore a useful model to describe the GWB produced by a realistic SMBHB population. That study models the common signal as a common uncorrelated red-noise (CURN) process and does not fit the HD correlations between pulsars.

In this article, {we test the TP on a GWB signal produced by a realistic population of SMBHBs, whose discrete nature produces both deviations from a PL and sky anisotropies.} {We infer the signal PSD using both a Bayesian estimate of the CURN and the frequentist per-frequency optimal statistic (PFOS)} \citep{gersb}, considering pair covariance and noise marginalization. The PFOS allows us not only to estimate the power and shape of the {HD-correlated common signal}, but also to {build maps of the GW power across the sky in individual frequency bins} through frequentist anisotropic methods \citep{ptr22}. This is useful to localize the loudest continuous gravitational waves (CGWs) and measure the anisotropic and isotropic signal-to-noise ratios (S/Ns) \citep{2026PhRvD.113j3031G,klm26}.

The TP was also suggested by \citet{2026PhRvD.113j3031G} as a model to be used in the Bayesian CURN search, since the choice of the spectral model assumed in the initial spectral estimate affects the result of the PFOS. In this work, we test this suggestion by fitting the data with the three models: TP, PL and FS. We show that the TP model allows us to recover, in the PFOS analysis, a {more precise estimate, with also modest improvements in accuracy,} of the injected power of a realistic SMBHB population than the PL or the FS. Moreover, it results in higher isotropic and anisotropic S/Ns.
In particular, among the frequency bins with significant anisotropy, the TP gives the largest fraction of bins in which the loudest binary is localized.
The paper is organized as follows. In Sec.~\ref{simulations} we present the simulated PTA, the injected SMBHB populations and the three models used to characterize the GWB spectrum. In Sec.~\ref{analysis} we describe the Bayesian and frequentist tools used to recover the GWB signal and to build the sky maps. In Sec.~\ref{results} we present the results of the analysis and the statistics comparing the three models over 100 realistic SMBHB populations. In Sec.~\ref{discussion} we summarize our findings and present our conclusions.

\section{Simulated PTA dataset}
\label{simulations}
The main goal of this project is to test the ability of the TP to model the GWB spectrum and compare its performance with that of the PL and FS. {We therefore require a dataset with a known injected signal to measure how well each model is able to recover it.} In this section, we describe the simulated PTA dataset and the realistic SMBHB populations that we injected. We then define the three models used in our analysis to describe the PSD of the GWB.

\subsection{Pulsar noise and realistic GWB injection}
\label{pta}

Our simulated PTA consists of 100 pulsars randomly distributed {across the sky}, at a distance of 1 kpc, and observed with a cadence of 14 days for 30 years. {We construct the array using the timing model of a real millisecond pulsar as a template.}
We simulate ToAs with the \texttt{libstempo} Python wrapper \citep{vallisneri} for the \texttt{Tempo2} pulsar timing package \citep{hem+2006, ehm+2006}, generating ToAs at the two observing radio frequencies of 1400 and 2200 MHz.

To simulate and analyze a large number of realizations {within a reasonable computational time}, we inject only white noise (WN) into the PTA, which we model as a stationary Gaussian process. Other stochastic {noise processes}, such as red noise (RN) and noise linked to variations of the dispersion measure \citep[DM, see e.g.][]{ vHLML09,Lentati+2013}, are not injected and will be included in future studies. WN has the property of having the same power at all frequencies and is described by two main parameters: an error scaling factor (EFAC, $E_F$), which multiplies the uncertainty of the ToA, and an error added in quadrature (EQUAD, $E_Q$), which describes additional contributions such as instrumental effects or pulse profile jitter \citep[and references therein]{vs18}. Thus, the ToA error is given by:
\begin{equation}
\label{eq:wn}
\sigma = \sqrt{E_F^2 \sigma_{\text{ToA}}^2 + E_Q^2},
\end{equation}
where $\sigma_{\text{ToA}}$ is the ToA uncertainty due to template fitting errors. We inject WN with the same statistical properties in each pulsar, setting $E_F=1$ and $E_Q=10^{-6}$ {s} as constant parameters. The full configuration of the simulated PTA is described in Table~\ref{tab:pta}.

\begin{table}
\caption{Configuration of the simulated PTA.}
\label{tab:pta}
\centering
\begin{tabular}{cc}
\hline
\hline
Settings & Simulated PTA \\
\hline
Number of pulsars          & 100      \\
$T$                        & 30 yrs   \\
Cadence                    & 14 days  \\
Pulsar distance            & 1 kpc  \\
Number of frequency bins   & 30       \\
EFAC                       & 1        \\
EQUAD                      & $10^{-6}$ \\
Observation frequencies    & 1400, 2200 MHz \\
\hline
\end{tabular}
\end{table}

We derive our SMBHB population from one of the models compiled by \citet{Sesana_2013} and \citet{rosado}. In those models, the distribution of MBHB per unit redshift, chirp mass and emitted GW frequency, $d^3N/(dz\,d\mathcal{M}\,d\ln f_r)$, is constructed from 
observed quantities such as  
the cosmic SMBHB merger rate, the SMBH--host-galaxy relation, the galaxy mass function, pair fraction, and SMBH mass accretion through galaxy mergers. For this paper we select a model that produces a GWB with amplitude $A_{\rm gw} \sim 2.4 \times 10^{-15}$ when modeling the GWB as a PL (see Section \ref{models} for more details), which is consistent with current EPTA observations \citep{EPTA_DR2}.
We draw 100 Monte Carlo realizations of the MBHB distribution, each consisting of  {$\approx 10^5$ binaries}. For each sampled binary we {extract the rest-frame masses}, redshift, observed GW frequency, and we add sky coordinates, inclination and polarization angles, randomly sampled from isotropic orientation and sky distributions.

\begin{figure}
\centering
\includegraphics[width=\linewidth]{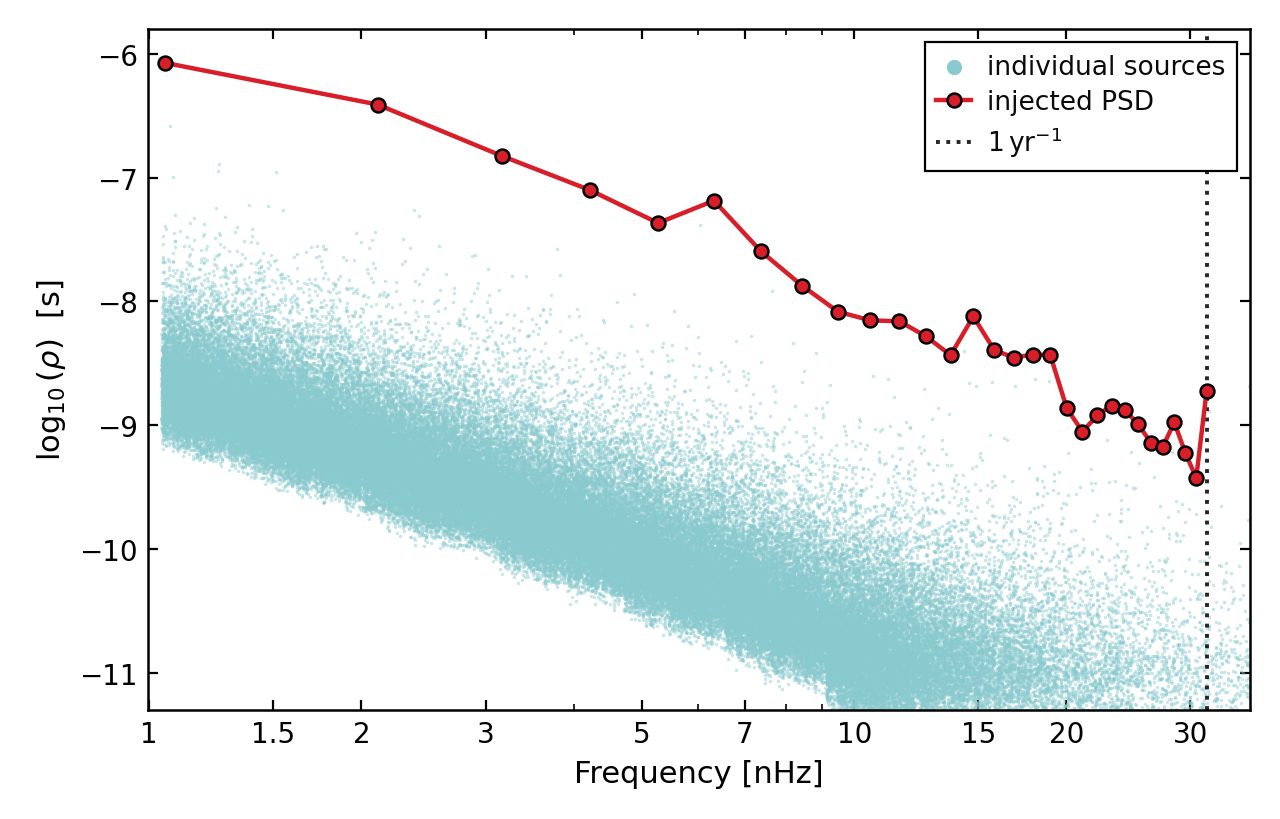}
\caption{PSD of one SMBHB population. Each light blue point is a single binary, and red dots
are the total injected PSD in each frequency bin, obtained by summing
the PSD contributions of all binaries that emit in each bin.\label{fig:injected_psd}}
\vspace*{-\baselineskip}
\end{figure}

We injected each SMBHB with eccentricity $e_0=0$ at formation, and we ignore any environmental coupling. This choice sets which deviations from a PL are present in the injected signal. The flattening at low frequencies due to  eccentricity and coupling with the environment are not tested in this work. Conversely, the finite number of sources in the population creates deviations from a PL, results in individually loud sources, and generates significant anisotropy, especially at high frequencies, all of which we aim to recover with the different spectral models. Fig.~\ref{fig:injected_psd} shows those features in an example of a SMBHB population; individual loud sources produce an excess of power in single frequency bins (as in the sixth one), while the population's discreteness results in an irregular PSD at higher frequencies.

{As noted above, WN is the only noise process injected into the dataset.} We use \texttt{discovery} \citep{2025zndo..17711453V} to draw one realization of Eq.~\eqref{eq:wn} from the measurement noise model of the array, {thereby obtaining the WN residuals for each pulsar.} Once the PTA is defined, we inject the GWB signal as an incoherent superposition of individual and independent sinusoidal waves produced by 
the SMBHBs drawn from the population as described above.
The pipeline used for the GWB injection takes as input a list of SMBHBs and their parameters, {and computes the residuals induced by each binary in the ToAs of every pulsar in the array, including both the Earth and pulsar terms and the evolution of the GW frequency.} Finally, the residuals of all binaries are summed in the time domain to produce the total GWB signal in each pulsar, {and this signal is added to the WN realization to obtain the residuals of the simulated dataset.}

To compare the PSD recovered in our analysis with the injected GWB, we compute an analytical estimate of the injected PSD in each frequency bin, obtained by summing the contributions of the individual binaries.
The strain of the GW signal emitted by a circular, GW-driven SMBHB is given by:
\begin{equation}
\label{eq:h_single}
h(f) = \sqrt{2 \left( \alpha^2 + \beta^2 \right)} \,
\frac{\left( G \mathcal{M} \right)^{5/3} \left( \pi f_r \right)^{2/3}}{c^4 r} ,
\end{equation}
where $\mathcal{M}$ is the chirp mass of the binary, $f_r = (1+z) f$ is the rest-frame GW frequency (twice the Keplerian frequency), $f$ is the observed GW frequency, $z$ is the redshift, $r$ is the comoving distance of the source, and $\alpha = 1 + \cos^2\iota$ {and} $\beta = -2\cos\iota$ are functions of the binary inclination angle $\iota$ \citep[see][for details]{rosado}.

Since we consider the GWB signal as a superposition of individual binaries, and the PTA observes for a finite time $T$ with a resolution of $\Delta f = 1/T$, we can evaluate the GWB spectrum in discrete frequency bins.
{For a frequency bin of width} $\Delta f$, the characteristic strain of a single binary is \citep{2010MNRAS.402.2308A}:
\begin{equation}
\label{eq:hc_single}
h_c^2(f) = \frac{h^2(f) \, f}{\Delta f} = h^2(f) \, f \, T .
\end{equation}
Thus, the contribution of a single binary to the PSD of the timing residuals is:
\begin{equation}
\label{eq:psd_single}
S(f) = \frac{ h^2(f)}{12 \pi^2 f^2} \, T .
\end{equation}
The injected PSD at a given frequency bin is obtained by summing Eq.~\eqref{eq:psd_single} over all binaries emitting in that bin. An example of the injected PSD and the overall population is shown in Fig.~\ref{fig:injected_psd}.

It is useful to comment on the frequency bin definition, because there are different ways to describe it. Both the Bayesian and frequentist analyses set the Fourier frequencies at $f_k = k/T$, with $k=1,\ldots,30$; the Fourier spacing, and hence the nominal frequency resolution, is $\Delta f=1/T$. We therefore define bins with these Fourier frequencies as their centers, which means that the edges of the $k$-th bin are at $(k-0.5)/T$ and $(k+0.5)/T$.
Given this discrete bin definition, a binary can emit at a frequency that is not exactly at the center of a bin.
Together with the finite observation time, this produces a leakage of power between neighboring bins, which is not accounted for by the diagonal covariance in Fourier space \citep{2025arXiv250613866C}. We do not correct for this effect in this project, and we defer to future work a treatment of the leakage due to the finite observation time and to the off-center frequencies of the emitting binaries.

\subsection{GWB models}
\label{models}

We describe the timing residuals of a pulsar as the difference between the measured ToAs and the ToAs predicted by the best-fit deterministic timing model:
\begin{equation}
\label{eq:res_def}
\delta t = t_{\rm ToA} - t_{\rm det}.
\end{equation}
{Thus, the residuals contain the stochastic processes not described by the deterministic timing model:}
\begin{equation}
\label{eq:residuals}
\delta t = M \vec{\epsilon} + F \vec{a} + \vec{n}_w,
\end{equation}
where $M$ is the linearized timing model design matrix and $\vec{\epsilon}$ are the linear perturbations of the timing model parameters, $F$ is the RN Fourier design matrix, whose columns are sine and cosine elements, $\vec{a}$ are the Fourier coefficients, which describe the intrinsic RN of the pulsar and the common signal produced by the GWB evaluated at the frequencies $f_k = k/T$, and $\vec{n}_w$ is the WN term of each pulsar. {As noted above, we inject only WN and the GWB signal; we do not include other stochastic processes, such as intrinsic pulsar RN or DM variations.}

Considering a population of circular SMBHBs whose evolution is driven by GW emission only, the GWB characteristic strain follows a PL \citep{phinney}:
\begin{equation}
\label{eq:hc}
h_c = A_{\rm gw} \left( \frac{f}{f_{\rm yr}} \right)^{-2/3},
\end{equation}
where $A_{\rm gw}$ is the amplitude at the frequency $f_{\rm yr} = 1\,{\rm yr}^{-1}$. The corresponding PSD of the residuals is:
\begin{equation}
\label{eq:pl}
P_{\rm pl}(f) = \frac{h_c^2(f)}{12\pi^2 f^3}\,\Delta f
= {\frac{A_{\rm gw}^2}{12\pi^2 f_{\rm yr}^{3}} \left( \frac{f}{f_{\rm yr}} \right)^{-\gamma_{\rm gw}}\Delta f} ,
\end{equation}
with $\gamma_{\rm gw}$ the spectral index, equal to $13/3$ for circular, GW-driven binaries. Most PTA searches use this model, defined by its amplitude $A_{\rm gw}$ and spectral index $\gamma_{\rm gw}$.

{As discussed in Section \ref{sec:intro}, a realistic population does not produce such a smooth spectrum.} The finite number of binaries that emit in each frequency bin, the coupling of the binaries with their environment, their eccentricity, and the presence of individually loud sources all produce deviations from the PSD of Eq.~\eqref{eq:pl}. Thus, we require a model that can capture these deviations.

Such a model is the FS, which describes the power in each frequency bin as an independent parameter:
\begin{equation}
\label{eq:fs}
P_{\rm fs}(f_i) = \frac{h_c^2(f_i)}{12\pi^2 f_i^3}\,\Delta f = \rho_i^2 ,
\end{equation}
where $\rho_i^2$ is the variance of the Fourier coefficients in the $i$-th bin. This model is described by one parameter per frequency bin; in our case, it has 30 parameters and is the most flexible of the three models considered here. However, this model does not impose any relation between different frequency bins; indeed, the FS fits the power in each frequency bin as a free parameter and makes no assumption on the shape of the spectrum. This freedom has a cost: each bin is constrained only by the data in that bin, and at high frequencies, where the WN dominates over the GWB, many bins are not constrained.

\begin{figure}
\centering
\includegraphics[width=0.8\linewidth]{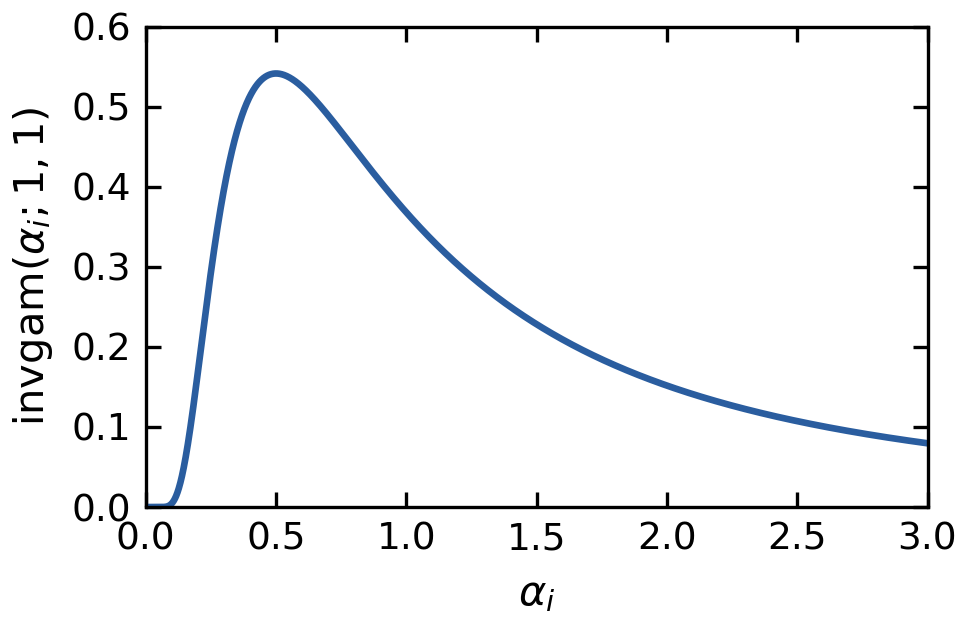}
\caption{Prior on the $\alpha_i$ parameters, described by an inverse gamma distribution.}
\label{fig:invgamma}
\end{figure}

The TP sits between these two models. The PL is kept as the underlying description of the spectrum, but every frequency bin is allowed to deviate from it by a multiplicative factor. We can write the PSD of the TP as:
\begin{equation}
\label{eq:tp}
P_{\rm tp}(f_i) = \alpha_i P_{\rm pl}(f_i)  \quad \alpha_i \sim {\rm invgam}(1,1)
\end{equation}
where $P_{\rm pl}(f_i)$ is the PL model described in Eq.~\eqref{eq:pl} and each $\alpha_i$ is a multiplicative factor that rescales the power in the $i$-th bin. We define the $\alpha_i$ prior as the inverse gamma distribution (Fig.~\ref{fig:invgamma}), which is a heavy-tailed distribution that allows large positive deviations from the PL, and therefore well suited to capture the excess power produced by individual loud sources. The probability density function of the inverse gamma distribution is given by:
\begin{equation}
\label{eq:invgam}
{\rm invgam}(\alpha_i,1,1) = \frac{1}{\Gamma(1)} \alpha_i^{-2}
\exp\left( -\frac{1}{\alpha_i} \right).
\end{equation}

The TP is therefore described by the two parameters of the underlying PL model plus an extra $\alpha_i$ parameter for each frequency bin. To illustrate the behavior of the $\alpha_i$ parameters, a value $\log \alpha_i = 0$ means that the PSD in that bin agrees with the underlying PL, while a value significantly different from zero {quantifies the deviation from it:} if $\log \alpha_i > 0$, the PSD measured is higher than the PL measurement, while if $\log \alpha_i < 0$, the measured PSD is lower.

The name of the model comes from the prior used on the Fourier coefficients. In standard RN models these coefficients have a Gaussian prior whose variance is set by the PL PSD; this defines a Gaussian process. In the TP, the variance is not fixed by the PL alone but is rescaled by $\alpha_i$, which is itself a random variable with an inverse gamma prior. If we marginalize the Gaussian over $\alpha_i$, we obtain a Student's t-distribution for the coefficients. The TP can therefore be viewed as a process in which marginalizing over the inverse-gamma variance scale gives Student's t-distributed Fourier coefficients.

The TP was not initially applied to model the GWB. It was introduced to model the RN of a single pulsar. {In particular, \citet{2019ApJ...880..116A} searched the NANOGrav 11-year dataset for possible continuous GWs from individual SMBHBs and found that} the then observed increase in the strain upper limit at $f_{\rm gw}=15$~nHz was generated by the pulsar PSR J0613$-$0200. {A dropout analysis found a much larger Bayes factor for that pulsar than for all the others.} Thus, the authors concluded that the increase in the upper limit was produced by an unmodeled feature in PSR J0613$-$0200 and not by a GW signal.

\citet{2020ApJ...900..102A} analyzed the same pulsar while searching for a continuous GW from the candidate SMBHB 3C 66B. The excess of power in PSR J0613$-$0200 RN made the sampling of the CGW frequency unreliable. Because the pulsar could not simply be removed from the array without losing sensitivity to the target search, its RN was described with a more flexible model, the TP spectrum, in which the Gaussian process prior was indeed generalized to a Student's t-distribution. The purpose was therefore to correctly model the excess of power in one frequency bin of that pulsar. This is possible because the heavy tail of the inverse gamma prior (Fig.~\ref{fig:invgamma}) allows the $\alpha_i$ of a single bin to assume large values.
A few years later, \citet{2023ApJ...951L...8A} applied this "fuzzy" model to {search for the GWB in the NANOGrav 15-year dataset.} Here the model was used to test how robust the recovered spectral index was against any deviations from a pure PL model.

Finally, \citet{2024ApJ...976..212S} studied the model systematically, comparing it with the PL and the FS on simulated datasets, and showed that the $\alpha_i$ parameters can be used to quantify where the spectrum deviates from a PL.
To summarize the history, the TP was initially applied to describe an excess of power present in a single pulsar's RN in a specific frequency bin on top of a smooth PL process. Then it was used to model the GWB, since we expect a similar signature in multiple frequency bins when the signal is sourced by a realistic SMBHB population.

\section{Analysis methods}
\label{analysis}
In this section we describe the Bayesian and frequentist techniques used to recover the GWB signal and to build sky maps of the GW power at different frequency bins.

\subsection{Bayesian methods}
\label{bayesian}

The standard Bayesian analysis of a PTA dataset is based on Bayes' theorem. Given the residuals $\delta t$ {defined above} and the parameter vector $\phi$, the posterior distribution is written as:
\begin{equation}
\label{eq:bayes}
p(\phi | \delta t) = \frac{p(\delta t | \phi) \, p(\phi)}{p(\delta t)},
\end{equation}
where $p(\delta t | \phi)$ is the likelihood function, $p(\phi)$ is the parameter prior, and $p(\delta t)$ is the evidence, which does not depend on $\phi$ and thus is considered as a normalization factor.

The likelihood function of the timing residuals, which we analytically marginalize over the timing model parameters \citep{vHLML09, vHV14}, is written as:
\begin{equation}
\label{eq:like}
p(\delta t | \phi) = \frac{\exp\left( -\frac{1}{2} \delta t^{T} \Sigma^{-1} \delta t \right)}
{\sqrt{(2\pi)^{n} \det \Sigma}},
\end{equation}
where $n$ is the total number of residuals, and $\Sigma$ is the covariance matrix of the array, which contains contributions from the stochastic processes present in the residuals. In our case, these processes are the WN and the GWB signal, which is a red noise process. The WN is uncorrelated between pulsars, and it contributes to the diagonal blocks of $\Sigma$, while the GWB contributes to both the diagonal and off-diagonal elements, because it is present in all pulsars' residuals and correlates them.

In fact, as already discussed, the GWB correlates the residuals of different pulsars depending on the angular separation of two pulsars in the sky. 
This correlation is the signature that separates a genuine GWB from other noise processes that can be present in each pulsar. For two pulsars $a$ and $b$, the expectation value of the  correlation takes the form:
\begin{equation}
\label{eq:hd}
\Gamma_{ab} = \frac{1}{2} - \frac{1-\cos \xi_{ab}}{4}
\left[ \frac{1}{2} - 3 \ln \left( \frac{1-\cos \xi_{ab}}{2} \right) \right]
+ \frac{1}{2} \delta_{ab},
\end{equation}
and is known as the HD correlation curve \citep{HD}. Here, $\xi_{ab}$ is the angular separation between the two pulsars and $\delta_{ab}$ is a Kronecker delta function.

In many PTA analyses, an initial search models the GWB as a common signal present in all pulsars but without spatial correlations between them. This is the CURN model in which, contrary to the HD model, the cross-pulsar blocks of $\Sigma$ are ignored. The CURN model is computationally cheaper than the HD model because the likelihood omits cross-pulsar terms. In this work we use the CURN model with the three spectral models described in Sec.\ref{models}, and we use the resulting chains as input for the frequentist analysis that we discussed in Sect.~\ref{freq}, which reconstructs the power under the HD hypothesis.

In recent PTA analyses, posterior distributions are typically estimated using Markov chain Monte Carlo (MCMC) sampling, implemented in \texttt{enterprise} \citep{Ellis_2019} and in \texttt{enterprise\_extensions} \citep{enterprise_extensions}. This approach is well tested and widely used in PTA analyses. It is, however, computationally heavy and may require a large number of samples to converge, which becomes a limitation for high-dimensional analyses with many pulsars and complex noise models.

For this reason, in this work we use the variational Bayesian inference approach \citep{2025PhRvL.135g1401V}, in which a normalizing flow is trained to approximate the posterior distributions. Since the likelihood is implemented in \texttt{JAX}, the whole computation can run on GPUs, which is what makes the procedure much faster than the MCMC sampler. The two methods differ in one important aspect: MCMC is asymptotically exact, meaning that the samples converge to the true posterior if the chain is run long enough, while variational inference returns an approximation, whose quality depends on the flexibility of the flow and the optimization.
In this work, we use the implementation provided by \texttt{discovery} \citep{2025zndo..17711453V}. We set uniform priors for the spectral index and log-uniform priors for the amplitude and $\rho_i$ parameters. Prior ranges are listed in Table~\ref{tab:priors}.

\begin{table}
\caption{Parameter prior ranges.}
\label{tab:priors}
\centering
\begin{tabular}{lcc}
\hline
\hline
Parameter & Prior type & Range \\
\hline
$\gamma_{\rm gw}$ & Uniform          & $[0, 7]$ \\
$A_{\rm gw}$      & log-Uniform      & $[10^{-18}, 10^{-12}]$ \\
$\rho_i$          & log-Uniform      & $[10^{-11}, 10^{-3}]$ \\
$\alpha_i$        & ${\rm invgam}(1,1)$ & $[10^{-4}, 10^{2}]$ \\
\hline
\end{tabular}
\tablefoot{$\gamma_{\rm gw}$ and $A_{\rm gw}$ are the PL parameters of Eq.~\eqref{eq:pl}, used also in the TP model.
$\rho_i$ are the FS parameters of Eq.~\eqref{eq:fs} and $\alpha_i$ are the parameters described in Eq.~\eqref{eq:tp} for the TP model.
Thus, in total, we have 2, 30, and 32 parameters for the PL, FS, and TP models respectively.}
\end{table}

\subsection{Frequentist GWB power reconstruction}
\label{freq}
{The Bayesian analysis described above provides the GWB spectrum under the CURN model.} {Our aim, however, is to measure the PSD associated with the HD correlations described in Eq.~\eqref{eq:hd}, as expected for a realistic SMBHB population.} It can be recovered with a Bayesian approach, but this is computationally expensive {because the likelihood must include the cross-pulsar terms.} For this reason, we use a frequentist approach that takes as input the CURN posterior and computes the HD PSD. In particular, in our analysis we use the per-frequency optimal statistic (PFOS) developed by \citet{gersb}\footnote{\url{https://github.com/GersbachKa/defiant}}.

In the "standard" optimal statistic (OS) \citep{Anholm} the cross-correlations between pulsar pairs are computed against a template of the GWB spectrum, which assumes a PL spectral shape. The PFOS removes this assumption by estimating the correlated power in each frequency bin separately, so that the shape of the spectrum is not forced to be a PL. This makes it well suited to a realistic GWB, whose spectrum is more complex than a simple PL.

In the PFOS approach, the pairwise estimator for pulsars $a$ and $b$ and its variance are defined in \citet{gersb, 2026PhRvD.113j3031G} as:
\begin{equation}
\label{eq:rho}
\rho_{ab}(f_j) = \frac{\textbf{X}_{a}^{T}\tilde{\phi}(f_j)\textbf{X}_b}
{\text{tr}\Big[ \textbf{Z}_a \tilde{\phi}(f_j)\textbf{Z}_b \Phi(f_j) \Big]},
\end{equation}
\begin{equation}
\label{eq:sigma_ab}
\sigma_{ab,0}(f_j)^{2} = \frac{\text{tr}\Big[ \textbf{Z}_a \tilde{\phi}(f_j) \textbf{Z}_b
\tilde{\phi}(f_j)\Big]}
{\text{tr}\Big[ \textbf{Z}_a \tilde{\phi}(f_j)\textbf{Z}_b \Phi(f_j) \Big]^{2}},
\end{equation}
where
\begin{equation}
\label{eq:XZ}
\textbf{X}_{a} = \textbf{F}_{a}^{T} \textbf{N}_{a}^{-1} \boldsymbol{\delta t}_{a} \qquad
\textbf{Z}_{a} = \textbf{F}_{a}^{T} \textbf{N}_{a}^{-1} \textbf{F}_{a}.
\end{equation}
Here $\textbf{F}_{a}$ is the Fourier design matrix of pulsar $a$, $\textbf{N}_{a}$ is its white noise covariance matrix, and $\boldsymbol{\delta t}_{a}$ is the vector of its residuals. The matrix $\tilde{\phi}(f_j)$ is a frequency selector matrix, and $\Phi(f_j)$ is the normalized Fourier domain covariance matrix.

Thus, we can write the PSD of the GWB and its uncertainty as:
\begin{equation}
\label{eq:S_hat}
\hat{S}(f_j) = \left[ \boldsymbol{\Gamma}^{T} \textbf{C}_{j}^{-1} \boldsymbol{\Gamma}
\right]^{-1} \boldsymbol{\Gamma}^{T} \textbf{C}_{j}^{-1} \boldsymbol{\rho}_j ,
\end{equation}
\begin{equation}
\label{eq:sigma_S}
\sigma_{\hat{S}(f_j),0} = \left[ \boldsymbol{\Gamma}^{T} \textbf{C}_{j}^{-1}
\boldsymbol{\Gamma} \right]^{-1/2} ,
\end{equation}
where $\textbf{C}_j$ is the covariance matrix of the pairwise estimators at the frequency $f_j$, $\boldsymbol{\Gamma}$ are the ORF factors for each pulsar pair (defined in Eq.~\eqref{eq:hd}), and $\boldsymbol{\rho}_j$ is the vector of the pairwise estimators $\rho_{ab}(f_j)$ of Eq.~\eqref{eq:rho}.

In this work, we paid particular attention to two aspects of the PFOS procedure that are not present in the original formulation of the optimal statistic. First, in our analysis we consider the covariance between pulsar pairs, thus we assume a non-diagonal covariance matrix $\textbf{C}_j$. The "standard" OS works in the weak-signal regime, in which the GWB has a negligible contribution to the total noise of each pulsar, and in that limit the covariance of the pairwise estimators can be taken as diagonal. This assumption is not valid once the GWB becomes the dominant red noise contribution in the lowest frequency bins. The off-diagonal elements grow with the strength of the signal, and dropping them returns an estimator whose variance is too small \citep{gersb}. Since in our analysis the injected background is loud in the lowest frequency bins, we replace the diagonal approximation with the full pulsar-pair covariance matrix defined as:
\begin{equation}
\label{eq:cov_pairs}
C_{ab,cd}(f_j) = \left\langle \rho_{ab}(f_j) \, \rho_{cd}(f_j) \right\rangle -
\left\langle \rho_{ab}(f_j) \right\rangle \left\langle \rho_{cd}(f_j) \right\rangle ,
\end{equation}
where $ab$ and $cd$ are two pulsar pairs. The diagonal approximation keeps only the elements with $ab = cd$, for which Eq.~\eqref{eq:cov_pairs} reduces to the variance $\sigma_{ab,0}(f_j)^2$ of Eq.~\eqref{eq:sigma_ab}.

The second aspect is the treatment of the uncertainty in the PFOS results. The estimator described in Eq.~\eqref {eq:rho} needs a guess value of the CURN signal to build $\Phi(f_j)$, and using only one value of the chain, such as the maximum likelihood, ignores the uncertainty of its posterior. Following the noise-marginalized approach of \citet{2018PhRvD..98d4003V}, we instead evaluate the statistic many times over draws from the posterior of the common signal, obtaining a distribution of the recovered PSD described in Eq.~\eqref{eq:S_hat} rather than a single number. In this work we use $500$ random draws, taken from the posterior obtained with the variational inference described in Sect.~\ref{bayesian}. Although it is called noise marginalization, in this work it is performed over the CURN parameters, since in the OS method the CURN is treated as a noise process.

Noise marginalization alone, however, is not enough.
Each draw returns an estimate of the power together with its own uncertainty $\sigma_{\hat{S}(f_j),0}$, and keeping only the value of $\hat{S}(f_j)$ while discarding the uncertainties makes the final distribution narrower than it really is. We therefore also apply the uncertainty sampling proposed by \citet{gersb}: every noise-marginalized estimate is replaced by a set of random draws from a Gaussian centered on that estimate $\hat{S}(f_j)$, with a standard deviation equal to its uncertainty $\sigma_{\hat{S}(f_j),0}$. The resulting distribution accounts for both the spread of the CURN chain and the uncertainty of the estimator itself. Thus, throughout this work, we use the PFOS with the pulsar-pair covariance (PC), the noise marginalization and the uncertainty sampling (NM). For brevity, we retain the name PFOS throughout, but here it denotes the PFOS evaluated with pulsar-pair covariance, noise marginalization, and uncertainty sampling.
We therefore run the whole procedure three times for each population, providing to the PFOS the CURN chains obtained by modeling the GWB with the PL, the FS, and the TP model, so that each spectral model produces its own reconstruction of the correlated power.

\subsection{Anisotropy and sky maps}
\label{skymaps}

As already discussed in Section \ref{sec:intro}, a GWB produced by a finite number of SMBHBs is not isotropic, because a small number of sources might dominate the power in some frequency bins. The distribution of the GW power in the sky can be built from the pairwise correlations of Eq.~\eqref{eq:rho} since the response of each pulsar pair depends on the direction of the GWB power.

Assuming that the uncertainties of the cross-correlations are Gaussian, the likelihood of the vector $\boldsymbol{\rho}_j$ of the $N_{\rm cc} = N_{\rm psr}(N_{\rm psr}-1)/2$ cross-correlations at the frequency $f_j$ is written as \citep{ptr22}:
\begin{equation}
\label{eq:map_like}
p(\boldsymbol{\rho}_j | \textbf{P}_j) = \frac{\exp\left[-\frac{1}{2}
(\boldsymbol{\rho}_j - \textbf{R} \textbf{P}_j)^{T} \textbf{C}_j^{-1}
(\boldsymbol{\rho}_j - \textbf{R} \textbf{P}_j)\right]}{\sqrt{\det(2\pi \textbf{C}_j)}},
\end{equation}
where $\textbf{P}_j$ is the GW power in each pixel of the sky at the frequency $f_j$, $\textbf{R}$ is the response matrix that maps the power of every pixel onto the response of each pulsar pair, and $\textbf{C}_j$ is the covariance matrix of the cross-correlations. We write the power of each pixel with the square-root spherical harmonic decomposition \citep{rut}:
\begin{equation}
\label{eq:sqrt_harmonic}
{P}(\hat{\Omega}, f_j)^{1/2} = \sum_{L=0}^{\infty} \sum_{M=-L}^{L} b_{LM}(f_j)
Y_{LM} (\hat{\Omega}),
\end{equation}
where $Y_{LM}$ are the real-valued spherical harmonics and $b_{LM}$ the coefficients of the decomposition. The number of multipoles that a PTA can resolve is limited by the size of the array, roughly as $l_{\rm max} \lesssim \sqrt{N_{\rm psr}}$ \citep{romano}, so the $100$ pulsars of our simulated PTA allow us to set $l_{\rm max}=L_{\rm max}=10$, and the sky is pixelized with ${\rm N_{side}}=8$. \footnote{The parameter ${\rm N_{side}}$ must be a power of 2 and satisfy ${\rm N_{side}} \leq \sqrt{N_{\rm psr}(N_{\rm psr}-1)/24}$.}
These settings imply a characteristic angular scale $\theta \sim 180\deg/l_{\rm max} \simeq 18\deg$. The advantage of this basis is that the reconstructed power is positive by definition, which is not guaranteed if the power is decomposed in spherical harmonics.

Maximizing the likelihood of Eq.~\eqref{eq:map_like} allows us to compute the coefficients $b_{LM}(f_j)$ and therefore the GWB power at each pixel. Using the \texttt{MAPS} pipeline \citep{ptr22}\footnote{\url{https://github.com/NihanPol/MAPS}}, the maximum likelihood is found numerically with the \texttt{LMFIT} package \citep{newville}. Following the procedure described in Sect.~\ref{freq}, we have $500$ noise-marginalized realizations of the cross-correlations in each frequency bin, and we then build sky maps from each of them computing the median power at each pixel to obtain the final sky map.
\label{res_bayes}
\begin{figure*}
\centering
\includegraphics[width=0.815\linewidth]{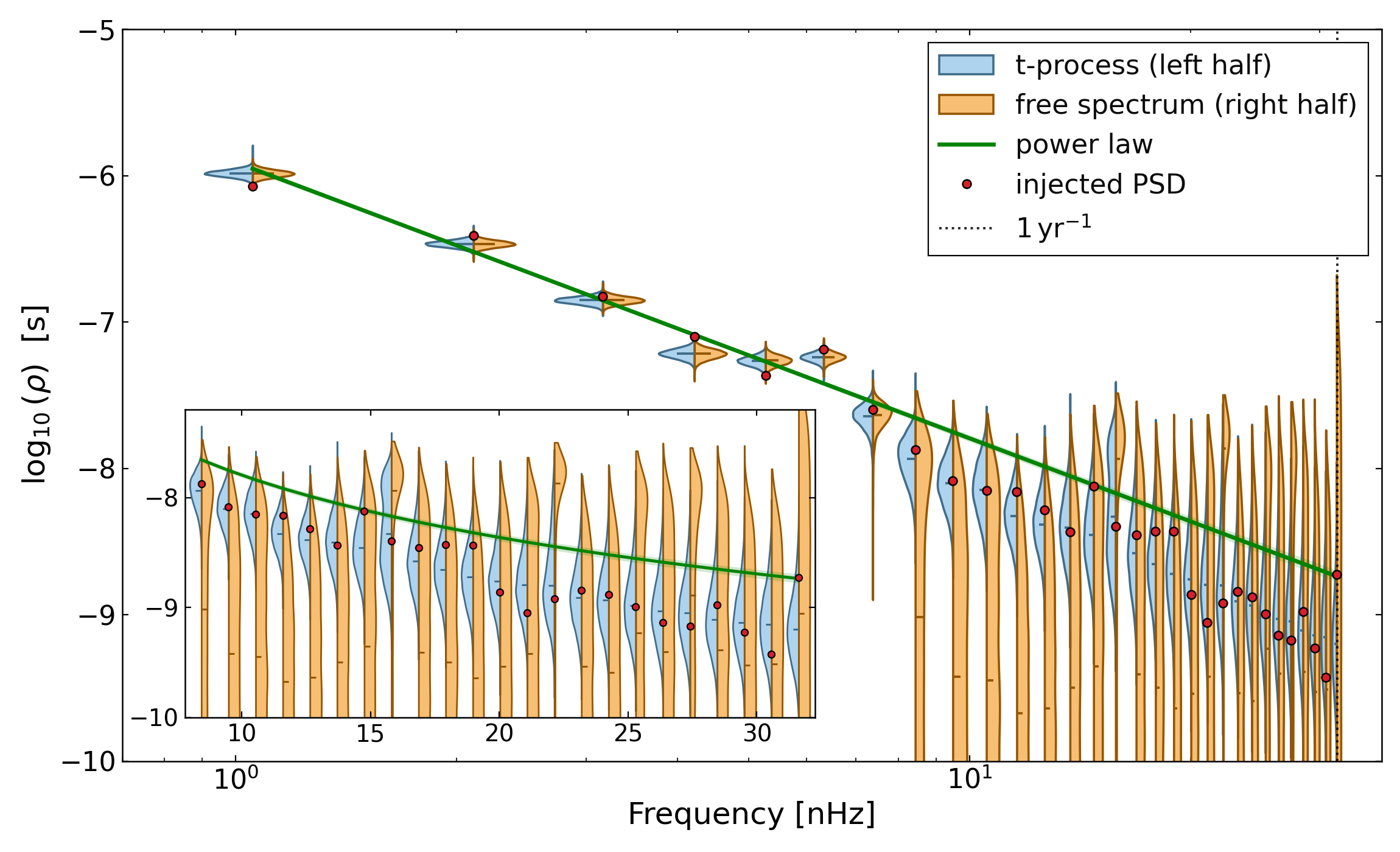}
\caption{GWB power spectral density recovered with the Bayesian tool for the three spectral models. The inset displays bins 8 to 30 on a linear x-axis, to show the results at high frequency more clearly.}
\label{fig:vi_spectra}
\end{figure*}
We can use the same likelihood to define a detection statistic. Computing the ratio between the maximum likelihood of two models gives the S/N values \citep{ptr22}:
\begin{equation}
\label{eq:snr}
\text{S/N}_j = \sqrt{2 \ln \left[ \frac{p(\boldsymbol{\rho} | \textbf{P}_{ML,1})}
{p(\boldsymbol{\rho} | \textbf{P}_{ML,2})}\right]}.
\end{equation}
We use this equation to compute two different S/N values. Taking the maximum likelihood ratio between the isotropic model and the noise only model gives the isotropic S/N, which measures the relative likelihood preference for an HD-correlated common signal over an uncorrelated common signal (CURN). On the other hand, the ratio of the maximum likelihood between the anisotropic and isotropic model gives the anisotropic S/N, which measures the relative likelihood preference for an anisotropic over an isotropic power distribution.
We thus compute the S/N at each frequency bin as the mean of the $500$ different S/N values computed from the different realizations. We evaluate these S/N values using the three spectral models, and we compare them to assess how the choice of spectral model affects the recovered anisotropic and HD-correlated power.

\section{Results}
\label{results}

In this section we present the results of the analysis. We first show the PSD recovered using the Bayesian approach, and then the GWB power computed by applying the PFOS analysis. We repeat the same procedure on 100 different populations, and we compare results using TP, FS, and PL models.

\subsection{Bayesian and frequentist results}

As a first step, we analyze each simulated dataset modeling the GWB as a TP, FS, and PL process, in all three cases as a CURN signal, computing the posterior distributions of the GWB parameters through the Bayesian analysis described in Sect.~\ref{bayesian}. Fig.~\ref{fig:vi_spectra} shows the spectra recovered in one SMBHB realization, together with the power that was injected in each frequency bin, as estimated from Eq.~\eqref{eq:psd_single}. {At low frequencies, where the GWB dominates, the TP and FS results are consistent with each other and with the injected PSD.} {At high frequencies, where WN begins to dominate, the FS results become uninformative, whereas the TP remains constrained and consistent with the injected PSD because it is tied to the underlying PL shape.} The PL model is not able to follow the injected PSD in each bin, since it is constrained to a PL shape that cannot capture any deviations from it.

These results are not completely new; \citet{2024ApJ...976..212S} analyzed simulated datasets based on the NANOGrav 15 yr array and found that the FS recovers the injected PSD at low frequency but loses information at the high frequencies where the WN dominates, while the TP keeps following the injected spectrum. We therefore do not repeat the characterization of the Bayesian posteriors and we do not build any statistic at this stage of the analysis.
We instead use these CURN chains only as input for the frequentist analysis, in which we compare the HD power recovered with the PFOS, using the three different spectral models, as described in Sect.~\ref{freq}. The goal is to establish which spectral model used in the Bayesian analysis improves the performance of the frequentist approach in computing the GWB power.

In Fig.~\ref{fig:pfos_reconstruction} we show the results of the PFOS analysis using the same population. {Using the TP chains, the PFOS yields more tightly constrained PSD values than when using the PL and FS Bayesian results, particularly in the first nine frequency bins.} {In bins 10--30, the three models give comparable results and the PFOS estimates are largely unconstrained.} In some of these bins, the PFOS estimates are not shown because the recovered power falls outside the plotted range, taking values close to or below zero. Negative estimates are expected since the PSD is computed through a linear minimization, which does not force the power to be positive. So, in bins where the GWB is weak compared with the WN, the estimate can fluctuate below zero. Bounding the power to positive values would require an iterative non-linear minimization, which is far less computationally efficient.

Given the intrinsic stochastic nature of the signal, the performance of a model cannot be assessed from the analysis of a single population. We therefore repeat the whole analysis on the 100 realizations described in Sect.~\ref{pta}, which allows us to quantitatively compare the three models with the statistics presented in the rest of this section.

\subsection{Statistics}
\label{res_stat}

The 100 datasets used in the analysis share the same pulsars, ToAs, and WN realization; the only difference is the SMBHB population that is injected in the PTA residuals. In our statistics, given two models $A$ and $B$ we compute the fraction of populations in which $A$ gives the better results, and the median relative gain ($G$) of results over the 100 populations is defined as:
\begin{equation}
\label{eq:gain}
G = {\rm median}_i \left[ 100 \,
\frac{X^{(B)}_i - X^{(A)}_i}{X^{(B)}_i} \right] ,
\end{equation}
where $X$ is the quantity being compared, and $i$ runs over the 100 populations. Eq.~\eqref{eq:gain} is written so that a model with a smaller quantity is better, as in the case of p-values described in Section~\ref{res_null} and for the precision and accuracy results in Section~\ref{res_acc}. For S/N results computed in Section~\ref{res_snr}, the sign at the numerator is reversed, so that a positive $G$ always indicates a gain for model $A$.

\subsubsection{Precision and accuracy}
\label{res_acc}
In our statistics, we call a bin informative when the recovered PFOS power exceeds its own uncertainty, thus $\hat{S}_k / \sigma_k > 1$. With this definition, we find that the median of this ratio over the 100 populations starts to be below one between the ninth and the tenth bin, around $10$~nHz. We therefore use in our model comparison the band between the 1st and 9th bin in which the PTA measures the GWB signal. Above the ninth bin, where the WN dominates, the recovered power falls below its own uncertainty in $77\%$ of the bins for the TP and the PL and in $99\%$ of them for the FS, so a comparison in this frequency range would be affected mostly by the noise fluctuations.

\begin{figure*}
\centering
\includegraphics[width=0.815\linewidth]{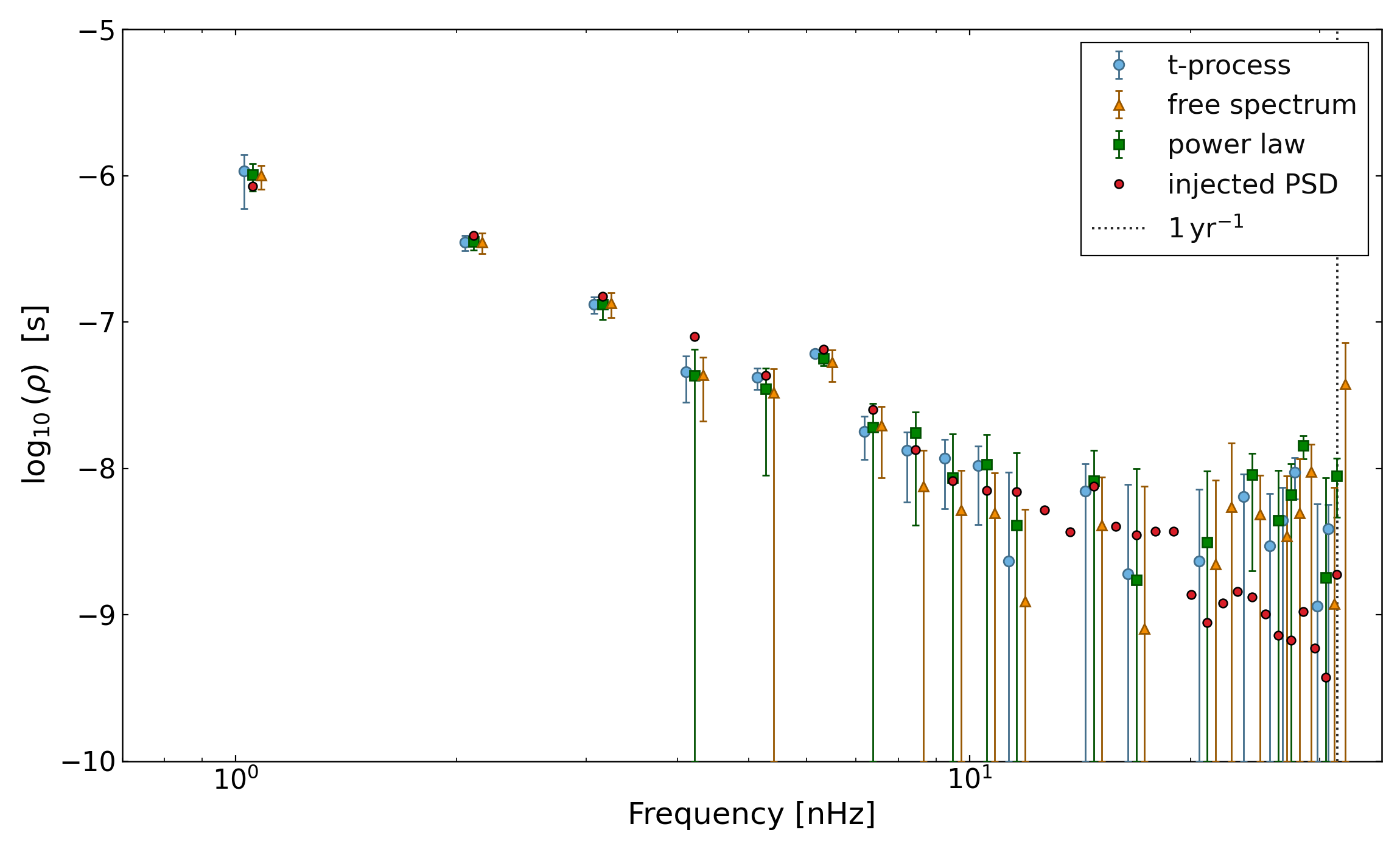}
\caption{GWB power spectral density computed with the PFOS tools using the same SMBHB realization as in
Fig.~\ref{fig:vi_spectra}. Each
model gives one estimate $\hat{S}(f_j)$ per frequency bin with its uncertainty, shown side by
side for comparison. The three estimates of the same bin are slightly shifted in frequency only
to make the figure clearer, all of them are computed at the same Fourier frequencies
$f_j = j/T$.}
\label{fig:pfos_reconstruction}
\end{figure*}

The first comparison that we perform is the PFOS results against the injected spectrum, which is known for all populations. We call this value ${\rm Inj}_k$ and we define in each bin:
\begin{equation}
\label{eq:tightclose}
{\rm T}_k = \frac{\sigma_k}{{\rm Inj}_k} , \qquad
{\rm C}_k = \frac{\left| S_k - {\rm Inj}_k \right|}{{\rm Inj}_k} .
\end{equation}
The first is the tightness (T) of the PFOS error bar, which measures the precision of the result. The second measures the relative absolute error, which we denote closeness (C). For each population, we then compute the median of these quantities over the bin range 1-9, obtaining three values (corresponding to the three signal models) for each population. Both tightness and closeness are normalized by the injected value, making the quantities dimensionless and more comparable across bins. Thus we count in how many populations the PFOS results are more precise and more accurate considering the TP model than using the FS and the PL ones, and we compute the median gain with Eq. \eqref{eq:gain}. Table~\ref{tab:acc} shows that the TP gives {substantially more precise} PFOS results, as we already noticed in Fig. \ref{fig:pfos_reconstruction}. Its accuracy is improved relative to the FS, while the difference relative to the PL is small.

\begin{table}
\caption{Precision and accuracy against the injected PSD}
\label{tab:acc}
\centering
\begin{tabular}{lccc}
\hline
\hline
Pair & Metric & Wins & Median gain \\
\hline
TP vs PL & T & 98/100 & $43.1\%$ \\
TP vs PL & C & 55/100 & $2.1\%$  \\
TP vs FS & T & 90/100 & $26.0\%$ \\
TP vs FS & C & 66/100 & $7.4\%$  \\
\hline
\end{tabular}
\tablefoot{
The wins count
the populations in which the TP model gives the smaller median value of the tightness or closeness.
The median gain is the median of the population-wise gains defined in Eq.~\eqref{eq:gain}. A positive value indicates a gain for the TP.}
\end{table}

\subsubsection{Isotropic and anisotropic S/N}
\label{res_snr}
In the previous section, we showed that using the TP chains, the PFOS {yields more tightly constrained estimates of the GWB power} in the frequency bins where the GWB dominates over the WN. {Beyond the recovered PSD, we can also ask how much of this power is associated with a GWB rather than noise, and how it is distributed across the sky.} To do that, for every population, every model and every frequency bin we evaluate Eq.~\eqref{eq:snr} on the PFOS results, obtaining the isotropic S/N, which compares the HD model with the absence of correlations between pulsars (CURN model), and the anisotropic S/N, which compares the anisotropic model with an isotropic one. We then sum in quadrature the S/N values at different frequency bins obtaining a single S/N value per population.

The results are listed in Table~\ref{tab:snr}. In the band 1-9, using all three models the S/N is above 30 for the isotropic case and above 40 for the anisotropic one, which indicates strong likelihood-ratio preferences for the correlated and anisotropic models in these simulations.
{In particular, the TP model gives higher anisotropic and isotropic S/N values in all 100 populations,} with a positive median gain over all 100 populations computed using Eq.~\eqref{eq:gain} (in which the sign of the numerator is reversed, because in this case a bigger value corresponds to a better model), indicating that, for these paired simulations, the TP-based posterior leads to larger values of both likelihood-ratio statistics than the FS- and PL-based posteriors.

In the table we also report the S/N values in the band 10-30 to show that in this range no information on the GWB can be extracted from the data. Summing in quadrature over the band, the median S/N is $5.0$ for the isotropic case and $15.2$ for the anisotropic one (using the TP chains), while the median S/N of a single bin is $0.5$ and $2.7$ respectively. An isotropic S/N below one corresponds to a weak likelihood-ratio preference for the HD-correlated model in a single bin; it does not by itself imply that the uncorrelated model is preferred. These anisotropic S/N values are expected to be higher than one even if only noise is present, as the anisotropic model has more free parameters than the isotropic one. For this reason they cannot be read as a significance, and we calibrate them against an isotropic null distribution described in Sect.~\ref{res_null}. The anisotropic S/N also decreases in the higher frequency bins because the correlations are noise dominated. In these bins the reconstructed structure reflects noise fluctuations rather than the distribution of GWB power.

This is worth stressing, because the injected population is intrinsically more anisotropic at high frequency, not less. Indeed the number of binaries that emit at each bin decreases as the frequency grows, so a smaller number of sources contribute to the total power, and the sky becomes more anisotropic. The measured anisotropic S/N does not follow this trend, and the reason is the level of the WN which dominates the signal in these bins. It is important to note that these results are median values taken over the 100 populations, thus this does not exclude the possibility that in a given population a single bin above the tenth is dominated by one loud binary. 
However, if we look at each bin in the range 1-9, the ratio between the median anisotropic and isotropic S/N increases with frequency, from $\sim 1.3$ to $\sim 2.4$, for all three spectral models. This means that, in the frequency range where the GWB dominates, the anisotropic model becomes more favored at higher frequency, as expected from the discreteness of the population.

\begin{table}[t]
\caption{Isotropic and anisotropic S/N}
\label{tab:snr}
\centering
\scriptsize
\setlength{\tabcolsep}{2pt}
\renewcommand{\arraystretch}{1.08}
\resizebox{\columnwidth}{!}{%
\begin{tabular}{@{}lccc cc cc@{}}
\hline
\hline
 & \multicolumn{3}{c}{Median S/N}
 & \multicolumn{2}{c}{vs.\ PL}
 & \multicolumn{2}{c}{vs.\ FS} \\
\cmidrule(lr){2-4}\cmidrule(lr){5-6}\cmidrule(lr){7-8}
Band & TP & PL & FS
& {\fontsize{8}{8}{TP $>$ PL}} & {\fontsize{8}{8}{Median gain}}
& {\fontsize{8}{8}{TP $>$ FS}} & {\fontsize{8}{8}{Median gain}} \\
\hline
\multicolumn{8}{c}{Isotropic S/N} \\
\hline
1--9   & 50.1 & 33.7 & 29.5 & 100/100 & $+45.6\%$ & 100/100 & $+64.9\%$ \\
10--30 &  5.0 &  4.9 &  5.0 &  97/100 &  $+2.7\%$ &  73/100 &  $+1.5\%$ \\
\hline
\multicolumn{8}{c}{Anisotropic S/N} \\
\hline
1--9   & 66.0 & 45.4 & 40.5 & 100/100 & $+46.0\%$ & 100/100 & $+65.4\%$ \\
10--30 & 15.2 & 14.7 & 14.9 &  97/100 &  $+2.7\%$ &  88/100 &  $+2.0\%$ \\
\hline
\end{tabular}%
}
\tablefoot{The first three columns are the median over the 100 populations of the S/N summed in
quadrature over the bins of the band. The count is the number of populations
in which the TP gives the larger S/N, and the gain beside it is the median computed using
Eq.~\eqref{eq:gain} over the 100 populations.}
\end{table}
\subsubsection{P-values and sky maps}
\label{res_null}

The anisotropic S/N cannot be read as a significance on its own. The null hypothesis is an isotropic GWB, and a single realization of an isotropic GWB is never perfectly isotropic, since it is produced by a finite number of sources whose waves can interfere. This is a form of realization (cosmic) variance, and it can make the anisotropic S/N nonzero even when the population has no preferred direction in the sky. The anisotropic S/N must therefore be calibrated against a distribution built from isotropic realizations.

The isotropic S/N would need a calibration as well. Such calibration can be performed with methods such as sky scrambles or phase shifts.
This is not necessary for our purposes, since we use the isotropic S/N only to compare the three spectral models applied to the same datasets, where pulsars, residuals and the injected population are the same.

To build the isotropic null distribution, we follow the procedure introduced in \citet{2026PhRvD.113j3031G}; we summarize the main steps here, leaving the reader to the original paper for a more detailed description. For each frequency bin, and for each of $500$ realizations, we:
\begin{enumerate}
\item add a gravitational wave in every pixel of the sky, drawing its amplitude from a Rayleigh distribution and its phase uniformly distributed in the range $[0, 2\pi)$. With this procedure, no direction is favored, but this particular realization is still not smooth;
\item propagate these waves in the PTA, with a response function containing pulsar and Earth terms, and combine them into the correlation of each pulsar pair, $\hat{\rho}_{ab,n}$;
\item scale the correlations by the PFOS power results and add the noise fluctuation,
\begin{equation}
\label{eq:null}
\vec{\rho}_{n,{\rm null}} = \mathcal{N}\left( \hat{S}_n \, \hat{\vec{\rho}}_{n},\;
\textbf{C}_{n,0} \right),
\end{equation}
where $\hat{S}_n$ is the PFOS estimate of the power in the bin $n$, $\hat{\vec{\rho}}_{n}$ are the correlations computed in the previous step and $\textbf{C}_{n,0}$ is the pair-independent covariance matrix of the PFOS in that bin. We refer to \citet{2026PhRvD.113j3031G} for the derivation of this equation;
\item apply the MAPS tool to $\vec{\rho}_{n,{\rm null}}$ and compute the anisotropic S/N using Eq. \eqref{eq:snr}.
\end{enumerate}
We stress that the values of $\hat{S}_n$ and the covariance are taken from the same posterior draw used by the noise marginalization in Section~\ref{freq}, so that the simulated sky and the measurement share the same PTA configuration, noise, and background strength. The only difference is that the simulation is isotropic by construction.

We evaluate the p-value only in bins where the recovered PSD is positive. Realizations of Eq.~\eqref{eq:null} are normalized by the power of the bin, so a value compatible with zero makes the normalized correlations diverge and return large S/N values that would bias the results. Such negative PSD results occur only where the WN dominates, therefore we consider these results as noise fluctuations. About twenty of the thirty bins in each population satisfy the positive-PSD condition.

{At the end of this procedure, we have $500$ S/N values drawn from the null distribution. The p-value in each bin is the fraction that exceed the measured S/N.} We then compute the p-value of the anisotropic S/N for all populations and for the three models. For each population we compute the median of the p-values over different frequency bins, providing one number per population and per model. We count the fraction of populations in which one model gives the smaller median p-value, and we measure the gain of these differences using Eq.~\eqref{eq:gain}. {We find that the TP gives a smaller median p-value than the FS in 91 of 100 populations and than the PL in 61 of 100.} The median gain is $22\%$ against the FS and only $4\%$ against the PL. These results show a clear reduction in median p-value relative to the FS, whereas the difference relative to the PL is {small} (61/100 populations; 4\% median gain).

\begin{figure*}
\centering
\includegraphics[width=0.67\linewidth]{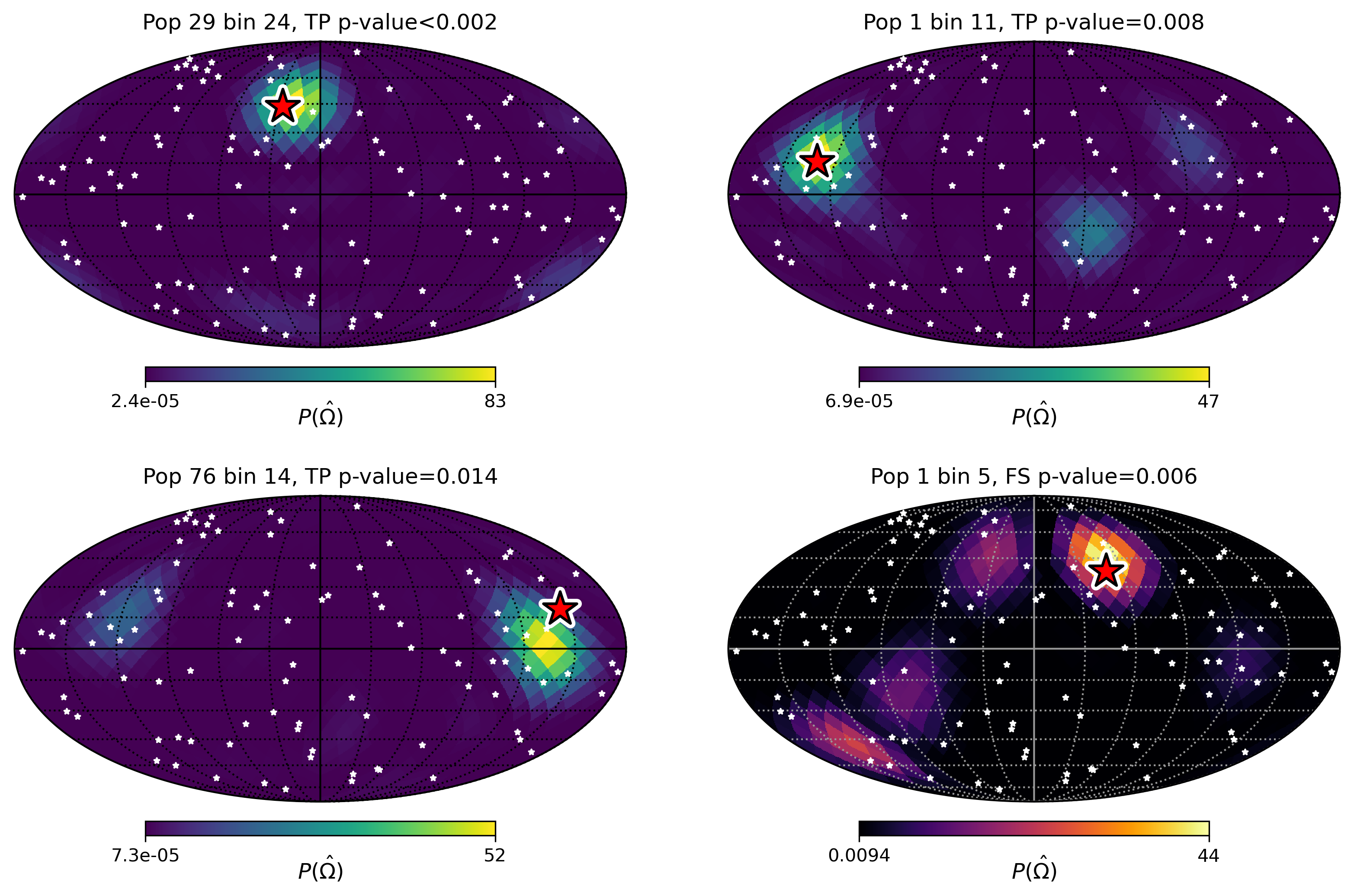}
\caption{Sky maps of the GW power computed following the procedure presented in Section~\ref{skymaps}.
The red star corresponds to the loudest CGW source emitting in the bin, and white stars are the pulsars of the PTA.
The top row and the bottom left panel are obtained using the TP model. The bottom right panel corresponds to the FS model.}
\label{fig:skymaps}
\end{figure*}

{We next determine how often the loudest source in a frequency bin can be localized.} {We adopt a threshold of $p<0.01$, selecting only bins with low p-values and therefore stronger evidence for anisotropy.} {We consider the loudest source localized when it lies within $18\deg$ of the maximum of the sky map (constructed as described in Section~\ref{skymaps}); this distance matches the angular resolution.} In our simulations, $14.3\%$ of the bins with $p<0.01$ using the TP model contain the loudest binary within this distance, against $10\%$ for the PL and $6.2\%$ for the FS. The list of these bins is shown in Table~\ref{tab:loc_bins}, where we report the offset between the maximum of the map and the loudest binary, together with the p-values for all three models. Within the selected $p<0.01$ bins, the fraction satisfying this localization criterion is higher for the TP than for the FS and PL models.
These results should not be read as a statement that one model builds better sky maps than another. Indeed, the sky maps are very similar among the three models; the main difference is the p-value assigned to a given bin. For illustration, Fig. \ref{fig:skymaps} shows four cases from Table \ref{tab:loc_bins} in which TP, FS, and PL are able to localize the loud CGW source.

\begin{table}
\caption{List of all frequency bins in which at least one of the three models has
$p<0.01$ and recovers the loudest binary within $18\deg$.}
\label{tab:loc_bins}
\centering
\setlength{\tabcolsep}{4pt}
\small
\begin{tabular}{cccccccc}
\hline
\hline
 &  & \multicolumn{3}{c}{offset [deg]} & \multicolumn{3}{c}{p-value} \\
\cmidrule(lr){3-5}\cmidrule(lr){6-8}
Pop. & Bin & TP & PL & FS & TP & PL & FS \\
\hline
\multicolumn{8}{c}{\textit{Recovered by the TP and by the PL}} \\
\hline
29 & 24 & $7.1$ & $7.1$ & $7.1$ & $<0.002$ & $0.002$ & $0.070$ \\
52 & 19 & $1.9$ & $1.9$ & $1.9$ & $0.004$ & $0.004$ & $0.142$ \\
69 & 5 & $15.0$ & $15.0$ & $15.0$ & $0.004$ & $0.004$ & $0.016$ \\
71 & 26 & $15.5$ & $15.5$ & $15.5$ & $0.004$ & $<0.002$ & $0.172$ \\
73 & 15 & $10.8$ & $10.8$ & $10.8$ & $<0.002$ & $<0.002$ & $0.120$ \\
98 & 17 & $16.5$ & $16.5$ & $16.5$ & $0.008$ & $0.002$ & $0.126$ \\
\hline
\multicolumn{8}{c}{\textit{Recovered by the TP only}} \\
\hline
\rowcolor{blue!10}1 & 11 & $4.0$ & $4.0$ & $4.0$ & $0.008$ & $0.014$ & $0.114$ \\
\rowcolor{blue!10}5 & 15 & $11.7$ & $11.7$ & $11.7$ & $0.006$ & $0.012$ & $0.034$ \\
\rowcolor{blue!10}7 & 15 & $16.0$ & $20.7$ & $16.0$ & $0.004$ & $0.004$ & $0.098$ \\
\rowcolor{blue!10}17 & 26 & $11.8$ & $11.8$ & $11.8$ & $0.008$ & $0.016$ & $0.120$ \\
\rowcolor{blue!10}58 & 10 & $15.4$ & $15.4$ & $15.4$ & $0.004$ & $0.018$ & $0.050$ \\
\rowcolor{blue!10}64 & 19 & $4.9$ & $4.9$ & $4.9$ & $0.008$ & $0.026$ & $0.088$ \\
\hline
\multicolumn{8}{c}{\textit{Recovered by the PL only}} \\
\hline
\rowcolor{red!10}76 & 14 & $17.9$ & $17.9$ & $17.9$ & $0.014$ & $0.008$ & $0.198$ \\
\hline
\multicolumn{8}{c}{\textit{Recovered by the PL and by the FS}} \\
\hline
\rowcolor{red!10}1 & 5 & $11.9$ & $11.9$ & $3.9$ & $0.038$ & $0.002$ & $0.006$ \\
\hline
\end{tabular}
\tablefoot{Blue rows correspond to bins that follow the 2 conditions ($p<0.01$ and CGW within $18\deg$) 
recovered by the TP and not by the PL and the FS, red rows correspond to bins recovered
by the PL and not by the TP. 
These two conditions are satisfied in 12, 8, and 1 bins using the TP, PL, and FS models respectively. 
In some cases we write $p<0.002$ because its value is lower than the resolution of the
null distribution, which is $1/500$.}
\end{table}

\section{Discussion and conclusions}
\label{discussion}

The GWB produced by a realistic population of SMBHBs is not expected to follow a PL. The finite number of binaries emitting in each frequency bin, coupling to the stellar environment, eccentricity, and individually loud sources can make the spectrum deviate from the smooth PL shape. Two models are commonly used in PTA searches: the PL and FS {models}. The former cannot follow these deviations, while the latter can, but constrains each frequency bin using only the information available in that bin; high-frequency bins can therefore become uninformative where WN dominates. The aim of this work was to assess whether a third model, the TP, provides a better description of the GWB.
The TP keeps a PL as the underlying description of the spectrum and {allows each frequency bin to deviate from it through a multiplicative factor with an inverse-gamma prior.} The model was first introduced to describe an excess of power in the RN of a single pulsar \citep{2019ApJ...880..116A, 2020ApJ...900..102A} and was later applied to the common signal \citep{2023ApJ...951L...8A, 2024ApJ...976..212S}.

To compare the three models we built a simulated PTA of 100 pulsars observed for 30 years with a cadence of 14 days, in which the WN is the only injected noise process. We then injected 100 realizations of a realistic SMBHB population in the time domain. Each of the 100 datasets was then analyzed three times, modeling the GWB as a PL, as a FS and as a TP. The analysis consists of three main steps. First, we estimated the CURN posterior distributions of each spectral model with variational inference. Second, we used these chains as the input of the PFOS, with the pulsar-pair covariance, noise marginalization, and uncertainty sampling, obtaining the HD correlated power in each frequency bin. Third, we built sky maps of the GW power at every frequency bin, and we calibrated their anisotropic S/N against a null distribution of isotropic realizations that accounts for cosmic variance. Since the 100 datasets share the same pulsars, ToAs, and WN realization, we compared the {results of the three models} population by population, counting the fraction of populations in which one model performs better than the others and measuring the median gain.

The Bayesian analysis confirmed what found by \citet{2024ApJ...976..212S}. At low frequency the TP and the FS agree with each other and with the injected PSD, while at high frequency the FS posteriors become uninformative, and the TP keeps following the injected spectrum because it is tied to an underlying PL. The PL, by construction, cannot reproduce any deviation from its own shape. The novel results of our work concern what happens when these chains are provided to the frequentist {PFOS} analysis.

We found that the per-frequency PSD reconstructed by the PFOS is more precise and accurate when the CURN process is modeled with the TP. In the 9 lowest frequency bins, where the GWB power is constrained by the array, {the uncertainties of the PFOS results obtained from the TP chains are} $43\%$ narrower than those obtained from the PL, and $26\%$ narrower than those obtained from the FS. The results are also closer to the injected PSD than the FS results, with a $7.4\%$ median gain, while relative to the PL the two reconstructions are at a comparable distance from the injected value, with a gain of $2.1\%$ for the TP.

We then computed the detection statistics, measuring how much of that power can be attributed to an HD correlated background, and how much of it is not uniformly distributed in the sky. Summed in quadrature over the first nine bins, the TP gives a larger value in all 100 populations, with a median gain of about $46\%$ over the PL and $65\%$ over the FS, {for both the anisotropic and isotropic S/N.} Above the tenth bin, where the WN dominates, this gain {decreases} to about $1.5$--$2\%$. Although systematic, this small gain in noise dominated bins does not indicate a real improvement of the TP over the other models.

{Finally, we computed the anisotropy p-values,} calibrating the anisotropic S/N against the isotropic null hypothesis. The TP gives a smaller median p-value than the FS, with a median gain of $22\%$. Against the PL the difference is much smaller, with a median gain of $4\%$, and we therefore consider these results as a marginal difference. Selecting the bins in which the anisotropy p-values are $p<0.01$, {the loudest injected CGW lies within $18\deg$ of the maximum of the sky map} in $14.3\%$ of the cases using the TP model, against $10\%$ with the PL and $6.2\%$ with the FS. This is not a statement about how well the sky maps are recovered; indeed different sky maps built with the three models agree with each other. What changes is the p-value that each model {assigns} to a similar sky map. The TP yields more selected bins in which the loudest binary satisfies the adopted localization criterion.

{These results show that the TP is a valid alternative to the PL and FS for modeling} the GWB produced by a realistic SMBHB population. It follows the deviations from a PL that such a population imprints on the spectrum, and, when it is used in the PFOS analysis, it gives the most precise reconstruction of the correlated power and the largest isotropic and anisotropic S/N of the three models. As the sensitivity of future PTAs improves and the shape of the GWB PSD has to be characterized more accurately, we therefore suggest that the TP be {considered alongside the PL and FS as a flexible model for characterizing the GWB spectrum}.

{Our analysis has several limitations that motivate future work.} First, we injected only WN into the datasets. Including intrinsic RN and DM variations is therefore an important next step to test whether the TP retains its advantages over the PL and FS models. Moreover, our simulated pulsars are evenly sampled, without any gaps and with almost the same time span, conditions in which the PFOS performs better \citep{gersb, 2026PhRvD.113j3031G}. In future work, we will therefore simulate more realistic pulsars, with unevenly sampled data, gaps, and different time spans.

Another limitation is that the injected SMBHB population used in this work consists of circular binaries without any environmental effects, so the only deviations from a PL that we have in our datasets are due to the discreteness of the population and to the individually loud sources. {Therefore, the comparison presented here assesses how well each model} {describes} the discreteness of a population, and not how well it follows a flatter PSD at low frequency. {Repeating the three-model comparison with eccentric binaries and environmental effects would also test the models at low frequencies.} 

Furthermore, we did not correct for the leakage of power between nearby frequency bins due to the finite observation time and the fact that the binaries do not emit exactly at the bin centers \citep{2025arXiv250613866C}.  As discussed above, the FS fits the power in each bin as a free parameter, while the TP keeps all the bins tied to a common PL, thus sharing information across bins. Whether this makes the TP less affected by leakage than the FS remains an open question that can be addressed with simulations designed specifically to quantify this effect. 

A further limitation is that the PFOS analyzes each frequency bin independently, as mentioned in Section~\ref{freq}. Modeling the CURN with the TP reduces the impact of this assumption, but it is still present in the PFOS analysis. Removing it would require a PFOS that analyzes all frequency bins simultaneously, including the covariances between them \citep{2026PhRvD.113j3031G}. Once these limitations are addressed, the next step will be to apply the pipeline to real PTA datasets.

\begin{acknowledgements}
We thank the B-Massive/Pingu group at Milano-Bicocca University for useful discussions and comments. B.E.M., A.S., and G.M.S. acknowledge the financial support provided under the European Union’s H2020 ERC Consolidator Grant “Binary Massive Black Hole Astrophysics” (B Massive, Grant Agreement: 818691) and the financial support provided under the European Union Advanced Grant “PINGU” (Grant Agreement: 101142079). Numerical calculations have been made possible through a CINECA-INFN agreement, providing access to resources on Galileo100/LEONARDO at CINECA. We acknowledge ISCRA for awarding this project access to the LEONARDO supercomputer, owned by the EuroHPC Joint Undertaking, hosted by CINECA (Italy), through the ISCRA class C project SBUBI. K.A.G. and S.R.T. acknowledge support from NSF CAREER \#2146016.
\end{acknowledgements}

\bibliographystyle{aa_edited}
\bibliography{draft}

@ARTICLE{ptr22,
       author = {{Pol}, Nihan and {Taylor}, Stephen R. and {Romano}, Joseph D.},
        title = "{Forecasting Pulsar Timing Array Sensitivity to Anisotropy in the Stochastic Gravitational Wave Background}",
      journal = {\apj},
         year = 2022,
        month = dec,
       volume = {940},
       number = {2},
          eid = {173},
        pages = {173},
          doi = {10.3847/1538-4357/ac9836},
archivePrefix = {arXiv},
       eprint = {2206.09936},
 primaryClass = {astro-ph.HE},
       adsurl = {https://ui.adsabs.harvard.edu/abs/2022ApJ...940..173P}
}

@ARTICLE{klm26,
       author = {{Konstandin}, Thomas and {Lemke}, Anna-Malin and {Mitridate}, Andrea and {Perboni}, Enrico},
        title = "{Prospects and limitations of PTAs anisotropy searches {\textemdash} the frequentist case}",
      journal = {\jcap},
         year = 2026,
        month = feb,
       volume = {2026},
       number = {2},
          eid = {084},
        pages = {084},
          doi = {10.1088/1475-7516/2026/02/084},
archivePrefix = {arXiv},
       eprint = {2509.07074},
 primaryClass = {astro-ph.CO},
       adsurl = {https://ui.adsabs.harvard.edu/abs/2026JCAP...02..084K}
}

@software{enterprise_extensions,
       author = {{Hazboun}, Jeff and {Aaron} and {Vigeland}, Sarah and {Taylor}, Stephen and {Baker}, Paul T. and {van Haasteren}, Rutger and {Sun}, Jerry and {NimaLaal} and {B{\'e}csy}, Bence and {achalumeau} and {mattsamson0203} and {gabefreedman} and {Meyers}, Pat and {Lamb}, William G and {Ellis}, Justin and {Vallisneri}, Michele and {siyuan-chen} and {Simon}, Joseph and {Pol}, Nihan and {Reardon}, Daniel and {Pennucci}, Tim and {Yu}, Wang-Wei and {Swiggum}, Joe and {kdolum} and {Drachler}, Brendan and {Shapiro-Albert}, Brent},
        title = "{nanograv/enterprise\_extensions: v3.0.3}",
         year = 2025,
        month = jun,
          eid = {10.5281/zenodo.15604870},
          doi = {10.5281/zenodo.15604870},
      version = {v3.0.3},
    publisher = {Zenodo},
       adsurl = {https://ui.adsabs.harvard.edu/abs/2025zndo..15604870H},
 howpublished = {\url{https://doi.org/10.5281/zenodo.15604870}}
}

@ARTICLE{2010MNRAS.402.2308A,
       author = {{Amaro-Seoane}, Pau and {Sesana}, Alberto and {Hoffman}, Loren and {Benacquista}, Matthew and {Eichhorn}, Christoph and {Makino}, Junichiro and {Spurzem}, Rainer},
        title = "{Triplets of supermassive black holes: astrophysics, gravitational waves and detection}",
      journal = {\mnras},
         year = 2010,
        month = mar,
       volume = {402},
       number = {4},
        pages = {2308-2320},
          doi = {10.1111/j.1365-2966.2009.16104.x},
archivePrefix = {arXiv},
       eprint = {0910.1587},
 primaryClass = {astro-ph.CO},
       adsurl = {https://ui.adsabs.harvard.edu/abs/2010MNRAS.402.2308A}
}

@ARTICLE{hem+2006,
       author = {{Hobbs}, G. and {Edwards}, R. and {Manchester}, R.},
        title = "{TEMPO2: a New Pulsar Timing Package}",
      journal = {Chinese Journal of Astronomy and Astrophysics Supplement},
         year = 2006,
        month = dec,
       volume = {6},
       number = {S2},
        pages = {189-192},
          doi = {10.1088/1009-9271/6/S2/35},
       adsurl = {https://ui.adsabs.harvard.edu/abs/2006ChJAS...6b.189H}
}

@ARTICLE{ehm+2006,
       author = {{Edwards}, R.~T. and {Hobbs}, G.~B. and {Manchester}, R.~N.},
        title = "{TEMPO2, a new pulsar timing package - II. The timing model and precision estimates}",
      journal = {\mnras},
         year = 2006,
        month = nov,
       volume = {372},
       number = {4},
        pages = {1549-1574},
          doi = {10.1111/j.1365-2966.2006.10870.x},
archivePrefix = {arXiv},
       eprint = {astro-ph/0607664},
 primaryClass = {astro-ph},
       adsurl = {https://ui.adsabs.harvard.edu/abs/2006MNRAS.372.1549E}
}

@ARTICLE{Lentati+2013,
       author = {{Lentati}, Lindley and {Alexander}, P. and {Hobson}, M.~P. and {Taylor}, S. and {Gair}, J. and {Balan}, S.~T. and {van Haasteren}, R.},
        title = "{Hyper-efficient model-independent Bayesian method for the analysis of pulsar timing data}",
      journal = {\prd},
         year = 2013,
        month = may,
       volume = {87},
       number = {10},
          eid = {104021},
        pages = {104021},
          doi = {10.1103/PhysRevD.87.104021},
archivePrefix = {arXiv},
       eprint = {1210.3578},
 primaryClass = {astro-ph.IM},
       adsurl = {https://ui.adsabs.harvard.edu/abs/2013PhRvD..87j4021L}
}

@ARTICLE{vHLML09,
       author = {{van Haasteren}, Rutger and {Levin}, Yuri and {McDonald}, Patrick and {Lu}, Tingting},
        title = "{On measuring the gravitational-wave background using Pulsar Timing Arrays}",
      journal = {\mnras},
         year = 2009,
        month = may,
       volume = {395},
       number = {2},
        pages = {1005-1014},
          doi = {10.1111/j.1365-2966.2009.14590.x},
archivePrefix = {arXiv},
       eprint = {0809.0791},
 primaryClass = {astro-ph},
       adsurl = {https://ui.adsabs.harvard.edu/abs/2009MNRAS.395.1005V}
}

@ARTICLE{vs18,
       author = {{Verbiest}, Joris P.~W. and {Shaifullah}, G.~M.},
        title = "{Measurement uncertainty in pulsar timing array experiments}",
      journal = {Classical and Quantum Gravity},
         year = 2018,
        month = jul,
       volume = {35},
       number = {13},
          eid = {133001},
        pages = {133001},
          doi = {10.1088/1361-6382/aac412},
       adsurl = {https://ui.adsabs.harvard.edu/abs/2018CQGra..35m3001V}
}

@ARTICLE{IPTA2023_Baker,
       author = {{Agazie}, G. and {Antoniadis}, J. and {Anumarlapudi}, A. and {Archibald}, A.~M. and {Arumugam}, P. and {Arumugam}, S. and {Arzoumanian}, Z. and {Askew}, J. and {Babak}, S. and {Bagchi}, M. and {Bailes}, M. and {Bak Nielsen}, A.-S. and {Baker}, P.~T. and {Bassa}, C.~G. and {Bathula}, A. and {B{\'e}csy}, B. and {Berthereau}, A. and {Bhat}, N.~D.~R. and {Blecha}, L. and {Bonetti}, M. and {Bortolas}, E. and {Brazier}, A. and {Brook}, P.~R. and {Burgay}, M. and {Burke-Spolaor}, S. and {Burnette}, R. and {Caballero}, R.~N. and {Cameron}, A. and {Case}, R. and {Chalumeau}, A. and {Champion}, D.~J. and {Chanlaridis}, S. and {Charisi}, M. and {Chatterjee}, S. and {Chatziioannou}, K. and {Cheeseboro}, B.~D. and {Chen}, S. and {Chen}, Z.-C. and {Cognard}, I. and {Cohen}, T. and {Coles}, W.~A. and {Cordes}, J.~M. and {Cornish}, N.~J. and {Crawford}, F. and {Cromartie}, H.~T. and {Crowter}, K. and {Cury{\l}o}, M. and {Cutler}, C.~J. and {Dai}, S. and {Dandapat}, S. and {Deb}, D. and {DeCesar}, M.~E. and {DeGan}, D. and {Demorest}, P.~B. and {Deng}, H. and {Desai}, S. and {Desvignes}, G. and {Dey}, L. and {Dhanda-Batra}, N. and {Di Marco}, V. and {Dolch}, T. and {Drachler}, B. and {Dwivedi}, C. and {Ellis}, J.~A. and {Falxa}, M. and {Feng}, Y. and {Ferdman}, R.~D. and {Ferrara}, E.~C. and {Fiore}, W. and {Fonseca}, E. and {Franchini}, A. and {Freedman}, G.~E. and {Gair}, J.~R. and {Garver-Daniels}, N. and {Gentile}, P.~A. and {Gersbach}, K.~A. and {Glaser}, J. and {Good}, D.~C. and {Goncharov}, B. and {Gopakumar}, A. and {Graikou}, E. and {Griessmeier}, J.-M. and {Guillemot}, L. and {G{\"u}ltekin}, K. and {Guo}, Y.~J. and {Gupta}, Y. and {Grunthal}, K. and {Hazboun}, J.~S. and {Hisano}, S. and {Hobbs}, G.~B. and {Hourihane}, S. and {Hu}, H. and {Iraci}, F. and {Islo}, K. and {Izquierdo-Villalba}, D. and {Jang}, J. and {Jawor}, J. and {Janssen}, G.~H. and {Jennings}, R.~J. and {Jessner}, A. and {Johnson}, A.~D. and {Jones}, M.~L. and {Joshi}, B.~C. and {Kaiser}, A.~R. and {Kaplan}, D.~L. and {Kapur}, A. and {Kareem}, F. and {Karuppusamy}, R. and {Keane}, E.~F. and {Keith}, M.~J. and {Kelley}, L.~Z. and {Kerr}, M. and {Key}, J.~S. and {Kharbanda}, D. and {Kikunaga}, T. and {Klein}, T.~C. and {Kolhe}, N. and {Kramer}, M. and {Krishnakumar}, M.~A. and {Kulkarni}, A. and {Laal}, N. and {Lackeos}, K. and {Lam}, M.~T. and {Lamb}, W.~G. and {Larsen}, B.~B. and {Lazio}, T.~J.~W. and {Lee}, K.~J. and {Levin}, Y. and {Lewandowska}, N. and {Littenberg}, T.~B. and {Liu}, K. and {Liu}, T. and {Liu}, Y. and {Lommen}, A. and {Lorimer}, D.~R. and {Lower}, M.~E. and {Luo}, J. and {Luo}, R. and {Lynch}, R.~S. and {Lyne}, A.~G. and {Ma}, C.-P. and {Maan}, Y. and {Madison}, D.~R. and {Main}, R.~A. and {Manchester}, R.~N. and {Mandow}, R. and {Mattson}, M.~A. and {McEwen}, A. and {McKee}, J.~W. and {McLaughlin}, M.~A. and {McMann}, N. and {Meyers}, B.~W. and {Meyers}, P.~M. and {Mickaliger}, M.~B. and {Miles}, M. and {Mingarelli}, C.~M.~F. and {Mitridate}, A. and {Natarajan}, P. and {Nathan}, R.~S. and {Ng}, C. and {Nice}, D.~J. and {Ni{\c{t}}u}, I.~C. and {Nobleson}, K. and {Ocker}, S.~K. and {Olum}, K.~D. and {Os{\l}owski}, S. and {Paladi}, A.~K. and {Parthasarathy}, A. and {Pennucci}, T.~T. and {Perera}, B.~B.~P. and {Perrodin}, D. and {Petiteau}, A. and {Petrov}, P. and {Pol}, N.~S. and {Porayko}, N.~K. and {Possenti}, A. and {Prabu}, T. and {Quelquejay Leclere}, H. and {Radovan}, H.~A. and {Rana}, P. and {Ransom}, S.~M. and {Ray}, P.~S. and {Reardon}, D.~J. and {Rogers}, A.~F. and {Romano}, J.~D. and {Russell}, C.~J. and {Samajdar}, A. and {Sanidas}, S.~A. and {Sardesai}, S.~C. and {Schmiedekamp}, A. and {Schmiedekamp}, C. and {Schmitz}, K. and {Schult}, L. and {Sesana}, A. and {Shaifullah}, G. and {Shannon}, R.~M. and {Shapiro-Albert}, B.~J. and {Siemens}, X. and {Simon}, J. and {Singha}, J.},
        title = "{Comparing Recent Pulsar Timing Array Results on the Nanohertz Stochastic Gravitational-wave Background}",
      journal = {\apj},
         year = 2024,
        month = may,
       volume = {966},
       number = {1},
          eid = {105},
        pages = {105},
          doi = {10.3847/1538-4357/ad36be},
archivePrefix = {arXiv},
       eprint = {2309.00693},
 primaryClass = {astro-ph.HE},
       adsurl = {https://ui.adsabs.harvard.edu/abs/2024ApJ...966..105A}
}

@ARTICLE{high_freq,
       author = {{Sesana}, A. and {Vecchio}, A. and {Colacino}, C.~N.},
        title = "{The stochastic gravitational-wave background from massive black hole binary systems: implications for observations with Pulsar Timing Arrays}",
      journal = {\mnras},
         year = 2008,
        month = oct,
       volume = {390},
       number = {1},
        pages = {192-209},
          doi = {10.1111/j.1365-2966.2008.13682.x},
archivePrefix = {arXiv},
       eprint = {0804.4476},
 primaryClass = {astro-ph},
       adsurl = {https://ui.adsabs.harvard.edu/abs/2008MNRAS.390..192S}
}

@ARTICLE{phinney,
       author = {{Phinney}, E.~S.},
        title = "{A Practical Theorem on Gravitational Wave Backgrounds}",
      journal = {arXiv e-prints},
         year = 2001,
        month = aug,
          eid = {astro-ph/0108028},
        pages = {astro-ph/0108028},
          doi = {10.48550/arXiv.astro-ph/0108028},
archivePrefix = {arXiv},
       eprint = {astro-ph/0108028},
 primaryClass = {astro-ph},
       adsurl = {https://ui.adsabs.harvard.edu/abs/2001astro.ph..8028P}
}

@software{newville,
       author = {{Newville}, Matt and {Otten}, Renee and {Nelson}, Andrew and {Ingargiola}, Antonino and {Stensitzki}, Till and {Allan}, Dan and {Fox}, Austin and {Carter}, Faustin and {Micha{\l}} and {Osborn}, Ray and {Pustakhod}, Dima and {Lneuhaus} and {Weigand}, Sebastian and {Glenn} and {Deil}, Christoph and {Mark} and {Hansen}, Allan L.~R. and {Pasquevich}, Gustavo and {Foks}, Leon and {Zobrist}, Nicholas and {Frost}, Oliver and {Beelen}, Alexandre and {Stuermer} and {Azelcer} and {Hannum}, Andrew and {Polloreno}, Anthony and {Hedegaard Nielsen}, Jens and {Caldwell}, Shane and {Almarza}, Anthony and {Persaud}, Arun},
        title = "{lmfit/lmfit-py: 1.0.3}",
         year = 2021,
        month = oct,
          eid = {10.5281/zenodo.5570790},
          doi = {10.5281/zenodo.5570790},
      version = {1.0.3},
    publisher = {Zenodo},
       adsurl = {https://ui.adsabs.harvard.edu/abs/2021zndo...5570790N},
 howpublished = {\url{https://doi.org/10.5281/zenodo.5570790}}
}

@ARTICLE{gersb,
       author = {{Gersbach}, Kyle A. and {Taylor}, Stephen R. and {Meyers}, Patrick M. and {Romano}, Joseph D.},
        title = "{Spatial and spectral characterization of the gravitational-wave background with the PTA optimal statistic}",
      journal = {\prd},
         year = 2025,
        month = jan,
       volume = {111},
       number = {2},
          eid = {023027},
        pages = {023027},
          doi = {10.1103/PhysRevD.111.023027},
archivePrefix = {arXiv},
       eprint = {2406.11954},
 primaryClass = {astro-ph.IM},
       adsurl = {https://ui.adsabs.harvard.edu/abs/2025PhRvD.111b3027G}
}

@ARTICLE{Sesana_2013,
       author = {{Sesana}, A.},
        title = "{Insights into the astrophysics of supermassive black hole binaries from pulsar timing observations}",
      journal = {Classical and Quantum Gravity},
         year = 2013,
        month = nov,
       volume = {30},
       number = {22},
          eid = {224014},
        pages = {224014},
          doi = {10.1088/0264-9381/30/22/224014},
archivePrefix = {arXiv},
       eprint = {1307.2600},
 primaryClass = {astro-ph.CO},
       adsurl = {https://ui.adsabs.harvard.edu/abs/2013CQGra..30v4014S}
}

@ARTICLE{taylor,
       author = {{Taylor}, S.~R. and {Mingarelli}, C.~M.~F. and {Gair}, J.~R. and {Sesana}, A. and {Theureau}, G. and {Babak}, S. and {Bassa}, C.~G. and {Brem}, P. and {Burgay}, M. and {Caballero}, R.~N. and {Champion}, D.~J. and {Cognard}, I. and {Desvignes}, G. and {Guillemot}, L. and {Hessels}, J.~W.~T. and {Janssen}, G.~H. and {Karuppusamy}, R. and {Kramer}, M. and {Lassus}, A. and {Lazarus}, P. and {Lentati}, L. and {Liu}, K. and {Os{\l}owski}, S. and {Perrodin}, D. and {Petiteau}, A. and {Possenti}, A. and {Purver}, M.~B. and {Rosado}, P.~A. and {Sanidas}, S.~A. and {Smits}, R. and {Stappers}, B. and {Tiburzi}, C. and {van Haasteren}, R. and {Vecchio}, A. and {Verbiest}, J.~P.~W. and {EPTA Collaboration}},
        title = "{Limits on Anisotropy in the Nanohertz Stochastic Gravitational Wave Background}",
      journal = {\prl},
         year = 2015,
        month = jul,
       volume = {115},
       number = {4},
          eid = {041101},
        pages = {041101},
          doi = {10.1103/PhysRevLett.115.041101},
archivePrefix = {arXiv},
       eprint = {1506.08817},
 primaryClass = {astro-ph.HE},
       adsurl = {https://ui.adsabs.harvard.edu/abs/2015PhRvL.115d1101T}
}

@ARTICLE{rut,
       author = {{Taylor}, Stephen R. and {van Haasteren}, Rutger and {Sesana}, Alberto},
        title = "{From bright binaries to bumpy backgrounds: Mapping realistic gravitational wave skies with pulsar-timing arrays}",
      journal = {\prd},
         year = 2020,
        month = oct,
       volume = {102},
       number = {8},
          eid = {084039},
        pages = {084039},
          doi = {10.1103/PhysRevD.102.084039},
archivePrefix = {arXiv},
       eprint = {2006.04810},
 primaryClass = {astro-ph.IM},
       adsurl = {https://ui.adsabs.harvard.edu/abs/2020PhRvD.102h4039T}
}

@ARTICLE{romano,
       author = {{Romano}, Joseph D. and {Cornish}, Neil. J.},
        title = "{Detection methods for stochastic gravitational-wave backgrounds: a unified treatment}",
      journal = {Living Reviews in Relativity},
         year = 2017,
        month = dec,
       volume = {20},
       number = {1},
          eid = {2},
        pages = {2},
          doi = {10.1007/s41114-017-0004-1},
archivePrefix = {arXiv},
       eprint = {1608.06889},
 primaryClass = {gr-qc},
       adsurl = {https://ui.adsabs.harvard.edu/abs/2017LRR....20....2R}
}

@ARTICLE{Anholm,
       author = {{Anholm}, Melissa and {Ballmer}, Stefan and {Creighton}, Jolien D.~E. and {Price}, Larry R. and {Siemens}, Xavier},
        title = "{Optimal strategies for gravitational wave stochastic background searches in pulsar timing data}",
      journal = {\prd},
         year = 2009,
        month = apr,
       volume = {79},
       number = {8},
          eid = {084030},
        pages = {084030},
          doi = {10.1103/PhysRevD.79.084030},
archivePrefix = {arXiv},
       eprint = {0809.0701},
 primaryClass = {gr-qc},
       adsurl = {https://ui.adsabs.harvard.edu/abs/2009PhRvD..79h4030A}
}

@software{vallisneri,
       author = {{Vallisneri}, Michele},
        title = "{libstempo: Python wrapper for Tempo2}",
 howpublished = {\href{https://ascl.net/2002.017}{Astrophysics Source Code Library, record ascl:2002.017}},
         year = 2020,
        month = feb,
          eid = {ascl:2002.017},
archivePrefix = {ascl},
       eprint = {2002.017},
       adsurl = {https://ui.adsabs.harvard.edu/abs/2020ascl.soft02017V}
}

@ARTICLE{rosado,
       author = {{Rosado}, Pablo A. and {Sesana}, Alberto and {Gair}, Jonathan},
        title = "{Expected properties of the first gravitational wave signal detected with pulsar timing arrays}",
      journal = {\mnras},
         year = 2015,
        month = aug,
       volume = {451},
       number = {3},
        pages = {2417-2433},
          doi = {10.1093/mnras/stv1098},
archivePrefix = {arXiv},
       eprint = {1503.04803},
 primaryClass = {astro-ph.HE},
       adsurl = {https://ui.adsabs.harvard.edu/abs/2015MNRAS.451.2417R}
}

@ARTICLE{EPTA_DR2,
       author = {{EPTA Collaboration} and {InPTA Collaboration} and {Antoniadis}, J. and {Arumugam}, P. and {Arumugam}, S. and {Babak}, S. and {Bagchi}, M. and {Bak Nielsen}, A.-S. and {Bassa}, C.~G. and {Bathula}, A. and {Berthereau}, A. and {Bonetti}, M. and {Bortolas}, E. and {Brook}, P.~R. and {Burgay}, M. and {Caballero}, R.~N. and {Chalumeau}, A. and {Champion}, D.~J. and {Chanlaridis}, S. and {Chen}, S. and {Cognard}, I. and {Dandapat}, S. and {Deb}, D. and {Desai}, S. and {Desvignes}, G. and {Dhanda-Batra}, N. and {Dwivedi}, C. and {Falxa}, M. and {Ferdman}, R.~D. and {Franchini}, A. and {Gair}, J.~R. and {Goncharov}, B. and {Gopakumar}, A. and {Graikou}, E. and {Grie{\ss}meier}, J.-M. and {Guillemot}, L. and {Guo}, Y.~J. and {Gupta}, Y. and {Hisano}, S. and {Hu}, H. and {Iraci}, F. and {Izquierdo-Villalba}, D. and {Jang}, J. and {Jawor}, J. and {Janssen}, G.~H. and {Jessner}, A. and {Joshi}, B.~C. and {Kareem}, F. and {Karuppusamy}, R. and {Keane}, E.~F. and {Keith}, M.~J. and {Kharbanda}, D. and {Kikunaga}, T. and {Kolhe}, N. and {Kramer}, M. and {Krishnakumar}, M.~A. and {Lackeos}, K. and {Lee}, K.~J. and {Liu}, K. and {Liu}, Y. and {Lyne}, A.~G. and {McKee}, J.~W. and {Maan}, Y. and {Main}, R.~A. and {Mickaliger}, M.~B. and {Ni{\c{t}}u}, I.~C. and {Nobleson}, K. and {Paladi}, A.~K. and {Parthasarathy}, A. and {Perera}, B.~B.~P. and {Perrodin}, D. and {Petiteau}, A. and {Porayko}, N.~K. and {Possenti}, A. and {Prabu}, T. and {Quelquejay Leclere}, H. and {Rana}, P. and {Samajdar}, A. and {Sanidas}, S.~A. and {Sesana}, A. and {Shaifullah}, G. and {Singha}, J. and {Speri}, L. and {Spiewak}, R. and {Srivastava}, A. and {Stappers}, B.~W. and {Surnis}, M. and {Susarla}, S.~C. and {Susobhanan}, A. and {Takahashi}, K. and {Tarafdar}, P. and {Theureau}, G. and {Tiburzi}, C. and {van der Wateren}, E. and {Vecchio}, A. and {Venkatraman Krishnan}, V. and {Verbiest}, J.~P.~W. and {Wang}, J. and {Wang}, L. and {Wu}, Z.},
        title = "{The second data release from the European Pulsar Timing Array. III. Search for gravitational wave signals}",
      journal = {\aap},
         year = 2023,
        month = oct,
       volume = {678},
          eid = {A50},
        pages = {A50},
          doi = {10.1051/0004-6361/202346844},
archivePrefix = {arXiv},
       eprint = {2306.16214},
 primaryClass = {astro-ph.HE},
       adsurl = {https://ui.adsabs.harvard.edu/abs/2023A&A...678A..50E}
}

@ARTICLE{PPTA_gwb,
       author = {{Reardon}, Daniel J. and {Zic}, Andrew and {Shannon}, Ryan M. and {Hobbs}, George B. and {Bailes}, Matthew and {Di Marco}, Valentina and {Kapur}, Agastya and {Rogers}, Axl F. and {Thrane}, Eric and {Askew}, Jacob and {Bhat}, N.~D. Ramesh and {Cameron}, Andrew and {Cury{\l}o}, Ma{\l}gorzata and {Coles}, William A. and {Dai}, Shi and {Goncharov}, Boris and {Kerr}, Matthew and {Kulkarni}, Atharva and {Levin}, Yuri and {Lower}, Marcus E. and {Manchester}, Richard N. and {Mandow}, Rami and {Miles}, Matthew T. and {Nathan}, Rowina S. and {Os{\l}owski}, Stefan and {Russell}, Christopher J. and {Spiewak}, Ren{\'e}e and {Zhang}, Songbo and {Zhu}, Xing-Jiang},
        title = "{Search for an Isotropic Gravitational-wave Background with the Parkes Pulsar Timing Array}",
      journal = {\apjl},
         year = 2023,
        month = jul,
       volume = {951},
       number = {1},
          eid = {L6},
        pages = {L6},
          doi = {10.3847/2041-8213/acdd02},
archivePrefix = {arXiv},
       eprint = {2306.16215},
 primaryClass = {astro-ph.HE},
       adsurl = {https://ui.adsabs.harvard.edu/abs/2023ApJ...951L...6R}
}

@ARTICLE{CPTA_gwb,
       author = {{Xu}, Heng and {Chen}, Siyuan and {Guo}, Yanjun and {Jiang}, Jinchen and {Wang}, Bojun and {Xu}, Jiangwei and {Xue}, Zihan and {Caballero}, R. Nicolas and {Yuan}, Jianping and {Xu}, Yonghua and {Wang}, Jingbo and {Hao}, Longfei and {Luo}, Jingtao and {Lee}, Kejia and {Han}, Jinlin and {Jiang}, Peng and {Shen}, Zhiqiang and {Wang}, Min and {Wang}, Na and {Xu}, Renxin and {Wu}, Xiangping and {Manchester}, Richard and {Qian}, Lei and {Guan}, Xin and {Huang}, Menglin and {Sun}, Chun and {Zhu}, Yan},
        title = "{Searching for the Nano-Hertz Stochastic Gravitational Wave Background with the Chinese Pulsar Timing Array Data Release I}",
      journal = {Research in Astronomy and Astrophysics},
         year = 2023,
        month = jul,
       volume = {23},
       number = {7},
          eid = {075024},
        pages = {075024},
          doi = {10.1088/1674-4527/acdfa5},
archivePrefix = {arXiv},
       eprint = {2306.16216},
 primaryClass = {astro-ph.HE},
       adsurl = {https://ui.adsabs.harvard.edu/abs/2023RAA....23g5024X}
}

@ARTICLE{meerkat,
       author = {{Miles}, Matthew T. and {Shannon}, Ryan M. and {Reardon}, Daniel J. and {Bailes}, Matthew and {Champion}, David J. and {Geyer}, Marisa and {Gitika}, Pratyasha and {Grunthal}, Kathrin and {Keith}, Michael J. and {Kramer}, Michael and {Kulkarni}, Atharva D. and {Nathan}, Rowina S. and {Parthasarathy}, Aditya and {Singha}, Jaikhomba and {Theureau}, Gilles and {Thrane}, Eric and {Abbate}, Federico and {Buchner}, Sarah and {Cameron}, Andrew D. and {Camilo}, Fernando and {Moreschi}, Beatrice E. and {Shaifullah}, Golam and {Shamohammadi}, Mohsen and {Possenti}, Andrea and {Krishnan}, Vivek Venkatraman},
        title = "{The MeerKAT Pulsar Timing Array: the first search for gravitational waves with the MeerKAT radio telescope}",
      journal = {\mnras},
         year = 2025,
        month = jan,
       volume = {536},
       number = {2},
        pages = {1489-1500},
          doi = {10.1093/mnras/stae2571},
archivePrefix = {arXiv},
       eprint = {2412.01153},
 primaryClass = {astro-ph.HE},
       adsurl = {https://ui.adsabs.harvard.edu/abs/2025MNRAS.536.1489M}
}

@ARTICLE{HD,
       author = {{Hellings}, R.~W. and {Downs}, G.~S.},
        title = "{Upper limits on the isotropic gravitational radiation background from pulsar timing analysis.}",
      journal = {\apjl},
         year = 1983,
        month = feb,
       volume = {265},
        pages = {L39-L42},
          doi = {10.1086/183954},
       adsurl = {https://ui.adsabs.harvard.edu/abs/1983ApJ...265L..39H}
}

@ARTICLE{Rajagopal_1995,
       author = {{Rajagopal}, Mohan and {Romani}, Roger W.},
        title = "{Ultra--Low-Frequency Gravitational Radiation from Massive Black Hole Binaries}",
      journal = {\apj},
         year = 1995,
        month = jun,
       volume = {446},
        pages = {543},
          doi = {10.1086/175813},
archivePrefix = {arXiv},
       eprint = {astro-ph/9412038},
 primaryClass = {astro-ph},
       adsurl = {https://ui.adsabs.harvard.edu/abs/1995ApJ...446..543R}
}

@ARTICLE{jb2003,
       author = {{Jaffe}, A.~H. and {Backer}, D.~C.},
        title = "{Gravitational Waves Probe the Coalescence Rate of Massive Black Hole Binaries}",
      journal = {\apj},
         year = 2003,
        month = feb,
       volume = {583},
       number = {2},
        pages = {616-631},
          doi = {10.1086/345443},
archivePrefix = {arXiv},
       eprint = {astro-ph/0210148},
 primaryClass = {astro-ph},
       adsurl = {https://ui.adsabs.harvard.edu/abs/2003ApJ...583..616J}
}

@ARTICLE{wl2003,
       author = {{Wyithe}, J. Stuart B. and {Loeb}, Abraham},
        title = "{Low-Frequency Gravitational Waves from Massive Black Hole Binaries: Predictions for LISA and Pulsar Timing Arrays}",
      journal = {\apj},
         year = 2003,
        month = jun,
       volume = {590},
       number = {2},
        pages = {691-706},
          doi = {10.1086/375187},
archivePrefix = {arXiv},
       eprint = {astro-ph/0211556},
 primaryClass = {astro-ph},
       adsurl = {https://ui.adsabs.harvard.edu/abs/2003ApJ...590..691W}
}

@ARTICLE{2000PhRvL..85.3761D,
       author = {{Damour}, Thibault and {Vilenkin}, Alexander},
        title = "{Gravitational Wave Bursts from Cosmic Strings}",
      journal = {\prl},
         year = 2000,
        month = oct,
       volume = {85},
       number = {18},
        pages = {3761-3764},
          doi = {10.1103/PhysRevLett.85.3761},
archivePrefix = {arXiv},
       eprint = {gr-qc/0004075},
 primaryClass = {gr-qc},
       adsurl = {https://ui.adsabs.harvard.edu/abs/2000PhRvL..85.3761D}
}

@ARTICLE{bartolo,
       author = {{Bartolo}, N. and {Matarrese}, S. and {Riotto}, A. and {V{\"a}ihk{\"o}nen}, A.},
        title = "{Maximal amount of gravitational waves in the curvaton scenario}",
      journal = {\prd},
         year = 2007,
        month = sep,
       volume = {76},
       number = {6},
          eid = {061302},
        pages = {061302},
          doi = {10.1103/PhysRevD.76.061302},
archivePrefix = {arXiv},
       eprint = {0705.4240},
 primaryClass = {astro-ph},
       adsurl = {https://ui.adsabs.harvard.edu/abs/2007PhRvD..76f1302B}
}

@ARTICLE{sorbo,
       author = {{Sorbo}, Lorenzo},
        title = "{Parity violation in the Cosmic Microwave Background from a pseudoscalar inflaton}",
      journal = {\jcap},
         year = 2011,
        month = jun,
       volume = {2011},
       number = {6},
          eid = {003},
        pages = {003},
          doi = {10.1088/1475-7516/2011/06/003},
archivePrefix = {arXiv},
       eprint = {1101.1525},
 primaryClass = {astro-ph.CO},
       adsurl = {https://ui.adsabs.harvard.edu/abs/2011JCAP...06..003S}
}

@ARTICLE{boyle,
       author = {{Boyle}, Latham A. and {Buonanno}, Alessandra},
        title = "{Relating gravitational wave constraints from primordial nucleosynthesis, pulsar timing, laser interferometers, and the CMB: Implications for the early universe}",
      journal = {\prd},
         year = 2008,
        month = aug,
       volume = {78},
       number = {4},
          eid = {043531},
        pages = {043531},
          doi = {10.1103/PhysRevD.78.043531},
archivePrefix = {arXiv},
       eprint = {0708.2279},
 primaryClass = {astro-ph},
       adsurl = {https://ui.adsabs.harvard.edu/abs/2008PhRvD..78d3531B}
}

@ARTICLE{hiramatsu,
       author = {{Hiramatsu}, Takashi and {Kawasaki}, Masahiro and {Saikawa}, Ken'ichi},
        title = "{On the estimation of gravitational wave spectrum from cosmic domain walls}",
      journal = {\jcap},
         year = 2014,
        month = feb,
       volume = {2014},
       number = {2},
          eid = {031},
        pages = {031},
          doi = {10.1088/1475-7516/2014/02/031},
archivePrefix = {arXiv},
       eprint = {1309.5001},
 primaryClass = {astro-ph.CO},
       adsurl = {https://ui.adsabs.harvard.edu/abs/2014JCAP...02..031H}
}

@ARTICLE{tomita,
       author = {{Tomita}, K.},
        title = "{Non-Linear Theory of Gravitational Instability in the Expanding Universe}",
      journal = {Progress of Theoretical Physics},
         year = 1967,
        month = may,
       volume = {37},
       number = {5},
        pages = {831-846},
          doi = {10.1143/PTP.37.831},
       adsurl = {https://ui.adsabs.harvard.edu/abs/1967PThPh..37..831T}
}

@ARTICLE{matarrese,
       author = {{Matarrese}, Sabino and {Pantano}, Ornella and {Saez}, Diego},
        title = "{General-relativistic approach to the nonlinear evolution of collisionless matter}",
      journal = {\prd},
         year = 1993,
        month = feb,
       volume = {47},
       number = {4},
        pages = {1311-1323},
          doi = {10.1103/PhysRevD.47.1311},
       adsurl = {https://ui.adsabs.harvard.edu/abs/1993PhRvD..47.1311M}
}

@ARTICLE{hindmarsh,
       author = {{Hindmarsh}, Mark and {Huber}, Stephan J. and {Rummukainen}, Kari and {Weir}, David J.},
        title = "{Gravitational Waves from the Sound of a First Order Phase Transition}",
      journal = {\prl},
         year = 2014,
        month = jan,
       volume = {112},
       number = {4},
          eid = {041301},
        pages = {041301},
          doi = {10.1103/PhysRevLett.112.041301},
archivePrefix = {arXiv},
       eprint = {1304.2433},
 primaryClass = {hep-ph},
       adsurl = {https://ui.adsabs.harvard.edu/abs/2014PhRvL.112d1301H}
}

@ARTICLE{kosowsky,
       author = {{Kosowsky}, Arthur and {Turner}, Michael S. and {Watkins}, Richard},
        title = "{Gravitational radiation from colliding vacuum bubbles}",
      journal = {\prd},
         year = 1992,
        month = jun,
       volume = {45},
       number = {12},
        pages = {4514-4535},
          doi = {10.1103/PhysRevD.45.4514},
       adsurl = {https://ui.adsabs.harvard.edu/abs/1992PhRvD..45.4514K}
}

@software{Ellis_2019,
       author = {{Ellis}, Justin A. and {Vallisneri}, Michele and {Taylor}, Stephen R. and {Baker}, Paul T.},
        title = "{ENTERPRISE: Enhanced Numerical Toolbox Enabling a Robust PulsaR Inference SuitE}",
 howpublished = {\href{https://ascl.net/1912.015}{Astrophysics Source Code Library, record ascl:1912.015}},
         year = 2019,
        month = dec,
          eid = {ascl:1912.015},
archivePrefix = {ascl},
       eprint = {1912.015},
       adsurl = {https://ui.adsabs.harvard.edu/abs/2019ascl.soft12015E}
}

@ARTICLE{ravi,
       author = {{Ravi}, V. and {Wyithe}, J.~S.~B. and {Shannon}, R.~M. and {Hobbs}, G. and {Manchester}, R.~N.},
        title = "{Binary supermassive black hole environments diminish the gravitational wave signal in the pulsar timing band}",
      journal = {\mnras},
         year = 2014,
        month = jul,
       volume = {442},
       number = {1},
        pages = {56-68},
          doi = {10.1093/mnras/stu779},
archivePrefix = {arXiv},
       eprint = {1404.5183},
 primaryClass = {astro-ph.CO},
       adsurl = {https://ui.adsabs.harvard.edu/abs/2014MNRAS.442...56R}
}

@ARTICLE{sampsoncornish,
       author = {{Sampson}, Laura and {Cornish}, Neil J. and {McWilliams}, Sean T.},
        title = "{Constraining the solution to the last parsec problem with pulsar timing}",
      journal = {\prd},
         year = 2015,
        month = apr,
       volume = {91},
       number = {8},
          eid = {084055},
        pages = {084055},
          doi = {10.1103/PhysRevD.91.084055},
archivePrefix = {arXiv},
       eprint = {1503.02662},
 primaryClass = {gr-qc},
       adsurl = {https://ui.adsabs.harvard.edu/abs/2015PhRvD..91h4055S}
}

@ARTICLE{Valtolina_2024,
       author = {{Valtolina}, Serena and {Shaifullah}, Golam and {Samajdar}, Anuradha and {Sesana}, Alberto},
        title = "{Testing strengths, limitations, and biases of current pulsar timing arrays' detection analyses on realistic data}",
      journal = {\aap},
         year = 2024,
        month = mar,
       volume = {683},
          eid = {A201},
        pages = {A201},
          doi = {10.1051/0004-6361/202348084},
archivePrefix = {arXiv},
       eprint = {2309.13117},
 primaryClass = {astro-ph.HE},
       adsurl = {https://ui.adsabs.harvard.edu/abs/2024A&A...683A.201V}
}

@ARTICLE{B_2023,
       author = {{B{\'e}csy}, Bence and {Cornish}, Neil J. and {Meyers}, Patrick M. and {Kelley}, Luke Zoltan and {Agazie}, Gabriella and {Anumarlapudi}, Akash and {Archibald}, Anne M. and {Arzoumanian}, Zaven and {Baker}, Paul T. and {Blecha}, Laura and {Brazier}, Adam and {Brook}, Paul R. and {Burke-Spolaor}, Sarah and {Casey-Clyde}, J. Andrew and {Charisi}, Maria and {Chatterjee}, Shami and {Chatziioannou}, Katerina and {Cohen}, Tyler and {Cordes}, James M. and {Crawford}, Fronefield and {Cromartie}, H. Thankful and {Crowter}, Kathryn and {DeCesar}, Megan E. and {Demorest}, Paul B. and {Dolch}, Timothy and {Ferrara}, Elizabeth C. and {Fiore}, William and {Fonseca}, Emmanuel and {Freedman}, Gabriel E. and {Garver-Daniels}, Nate and {Gentile}, Peter A. and {Glaser}, Joseph and {Good}, Deborah C. and {G{\"u}ltekin}, Kayhan and {Hazboun}, Jeffrey S. and {Hourihane}, Sophie and {Jennings}, Ross J. and {Johnson}, Aaron D. and {Jones}, Megan L. and {Kaiser}, Andrew R. and {Kaplan}, David L. and {Kerr}, Matthew and {Key}, Joey S. and {Laal}, Nima and {Lam}, Michael T. and {Lamb}, William G. and {W. Lazio}, T. Joseph and {Lewandowska}, Natalia and {Littenberg}, Tyson B. and {Liu}, Tingting and {Lorimer}, Duncan R. and {Luo}, Jing and {Lynch}, Ryan S. and {Ma}, Chung-Pei and {Madison}, Dustin R. and {McEwen}, Alexander and {McKee}, James W. and {McLaughlin}, Maura A. and {McMann}, Natasha and {Meyers}, Bradley W. and {Mingarelli}, Chiara M.~F. and {Mitridate}, Andrea and {Ng}, Cherry and {Nice}, David J. and {Ocker}, Stella Koch and {Olum}, Ken D. and {Pennucci}, Timothy T. and {Perera}, Benetge B.~P. and {Pol}, Nihan S. and {Radovan}, Henri A. and {Ransom}, Scott M. and {Ray}, Paul S. and {Romano}, Joseph D. and {Sardesai}, Shashwat C. and {Schmiedekamp}, Ann and {Schmiedekamp}, Carl and {Schmitz}, Kai and {Shapiro-Albert}, Brent J. and {Siemens}, Xavier and {Simon}, Joseph and {Siwek}, Magdalena S. and {Sosa Fiscella}, Sophia V. and {Stairs}, Ingrid H. and {Stinebring}, Daniel R. and {Stovall}, Kevin and {Susobhanan}, Abhimanyu and {Swiggum}, Joseph K. and {Taylor}, Stephen R. and {Turner}, Jacob E. and {Unal}, Caner and {Vallisneri}, Michele and {van Haasteren}, Rutger and {Vigeland}, Sarah J. and {Wahl}, Haley M. and {Witt}, Caitlin A. and {Young}, Olivia},
        title = "{How to Detect an Astrophysical Nanohertz Gravitational Wave Background}",
      journal = {\apj},
         year = 2023,
        month = dec,
       volume = {959},
       number = {1},
          eid = {9},
        pages = {9},
          doi = {10.3847/1538-4357/ad09e4},
archivePrefix = {arXiv},
       eprint = {2309.04443},
 primaryClass = {gr-qc},
       adsurl = {https://ui.adsabs.harvard.edu/abs/2023ApJ...959....9B}
}

@ARTICLE{sesanavecchiovolonteri,
       author = {{Sesana}, A. and {Vecchio}, A. and {Volonteri}, M.},
        title = "{Gravitational waves from resolvable massive black hole binary systems and observations with Pulsar Timing Arrays}",
      journal = {\mnras},
         year = 2009,
        month = apr,
       volume = {394},
       number = {4},
        pages = {2255-2265},
          doi = {10.1111/j.1365-2966.2009.14499.x},
archivePrefix = {arXiv},
       eprint = {0809.3412},
 primaryClass = {astro-ph},
       adsurl = {https://ui.adsabs.harvard.edu/abs/2009MNRAS.394.2255S}
}

@ARTICLE{ravi2012,
       author = {{Ravi}, V. and {Wyithe}, J.~S.~B. and {Hobbs}, G. and {Shannon}, R.~M. and {Manchester}, R.~N. and {Yardley}, D.~R.~B. and {Keith}, M.~J.},
        title = "{Does a ``Stochastic'' Background of Gravitational Waves Exist in the Pulsar Timing Band?}",
      journal = {\apj},
         year = 2012,
        month = dec,
       volume = {761},
       number = {2},
          eid = {84},
        pages = {84},
          doi = {10.1088/0004-637X/761/2/84},
archivePrefix = {arXiv},
       eprint = {1210.3854},
 primaryClass = {astro-ph.CO},
       adsurl = {https://ui.adsabs.harvard.edu/abs/2012ApJ...761...84R}
}

@ARTICLE{kelley2017,
       author = {{Kelley}, Luke Zoltan and {Blecha}, Laura and {Hernquist}, Lars and {Sesana}, Alberto and {Taylor}, Stephen R.},
        title = "{The gravitational wave background from massive black hole binaries in Illustris: spectral features and time to detection with pulsar timing arrays}",
      journal = {\mnras},
         year = 2017,
        month = nov,
       volume = {471},
       number = {4},
        pages = {4508-4526},
          doi = {10.1093/mnras/stx1638},
archivePrefix = {arXiv},
       eprint = {1702.02180},
 primaryClass = {astro-ph.HE},
       adsurl = {https://ui.adsabs.harvard.edu/abs/2017MNRAS.471.4508K}
}

@ARTICLE{b_2025,
       author = {{B{\'e}csy}, Bence and {Cornish}, Neil J. and {Petrov}, Polina and {Siemens}, Xavier and {Taylor}, Stephen R. and {Vigeland}, Sarah J. and {Witt}, Caitlin A.},
        title = "{Towards robust gravitational wave detections from individual supermassive black hole binaries}",
      journal = {Classical and Quantum Gravity},
         year = 2025,
        month = sep,
       volume = {42},
       number = {17},
          eid = {175016},
        pages = {175016},
          doi = {10.1088/1361-6382/adfd36},
archivePrefix = {arXiv},
       eprint = {2502.18114},
 primaryClass = {gr-qc},
       adsurl = {https://ui.adsabs.harvard.edu/abs/2025CQGra..42q5016B}
}

@ARTICLE{Ferranti_2025,
       author = {{Ferranti}, Irene and {Shaifullah}, Golam and {Chalumeau}, Aurelien and {Sesana}, Alberto},
        title = "{Separating deterministic and stochastic gravitational wave signals in realistic pulsar timing array datasets}",
      journal = {\aap},
         year = 2025,
        month = feb,
       volume = {694},
          eid = {A194},
        pages = {A194},
          doi = {10.1051/0004-6361/202451809},
archivePrefix = {arXiv},
       eprint = {2407.21105},
 primaryClass = {astro-ph.HE},
       adsurl = {https://ui.adsabs.harvard.edu/abs/2025A&A...694A.194F}
}

@ARTICLE{Gair_2014,
       author = {{Gair}, Jonathan and {Romano}, Joseph D. and {Taylor}, Stephen and {Mingarelli}, Chiara M.~F.},
        title = "{Mapping gravitational-wave backgrounds using methods from CMB analysis: Application to pulsar timing arrays}",
      journal = {\prd},
         year = 2014,
        month = oct,
       volume = {90},
       number = {8},
          eid = {082001},
        pages = {082001},
          doi = {10.1103/PhysRevD.90.082001},
archivePrefix = {arXiv},
       eprint = {1406.4664},
 primaryClass = {gr-qc},
       adsurl = {https://ui.adsabs.harvard.edu/abs/2014PhRvD..90h2001G}
}

@ARTICLE{Mingarelli_2013,
       author = {{Mingarelli}, C.~M.~F. and {Sidery}, T. and {Mandel}, I. and {Vecchio}, A.},
        title = "{Characterizing gravitational wave stochastic background anisotropy with pulsar timing arrays}",
      journal = {\prd},
         year = 2013,
        month = sep,
       volume = {88},
       number = {6},
          eid = {062005},
        pages = {062005},
          doi = {10.1103/PhysRevD.88.062005},
archivePrefix = {arXiv},
       eprint = {1306.5394},
 primaryClass = {astro-ph.HE},
       adsurl = {https://ui.adsabs.harvard.edu/abs/2013PhRvD..88f2005M}
}

@ARTICLE{Olmez_2012,
       author = {{{\"O}lmez}, S. and {Mandic}, V. and {Siemens}, X.},
        title = "{Anisotropies in the gravitational-wave stochastic background}",
      journal = {\jcap},
         year = 2012,
        month = jul,
       volume = {2012},
       number = {7},
          eid = {009},
        pages = {009},
          doi = {10.1088/1475-7516/2012/07/009},
archivePrefix = {arXiv},
       eprint = {1106.5555},
 primaryClass = {astro-ph.CO},
       adsurl = {https://ui.adsabs.harvard.edu/abs/2012JCAP...07..009O}
}

@ARTICLE{Mingarelli_2017,
       author = {{Mingarelli}, Chiara M.~F. and {Lazio}, T. Joseph W. and {Sesana}, Alberto and {Greene}, Jenny E. and {Ellis}, Justin A. and {Ma}, Chung-Pei and {Croft}, Steve and {Burke-Spolaor}, Sarah and {Taylor}, Stephen R.},
        title = "{The local nanohertz gravitational-wave landscape from supermassive black hole binaries}",
      journal = {Nature Astronomy},
         year = 2017,
        month = nov,
       volume = {1},
        pages = {886-892},
          doi = {10.1038/s41550-017-0299-6},
archivePrefix = {arXiv},
       eprint = {1708.03491},
 primaryClass = {astro-ph.GA},
       adsurl = {https://ui.adsabs.harvard.edu/abs/2017NatAs...1..886M}
}

@ARTICLE{Taylor_2013,
       author = {{Taylor}, Stephen R. and {Gair}, Jonathan R.},
        title = "{Searching for anisotropic gravitational-wave backgrounds using pulsar timing arrays}",
      journal = {\prd},
         year = 2013,
        month = oct,
       volume = {88},
       number = {8},
          eid = {084001},
        pages = {084001},
          doi = {10.1103/PhysRevD.88.084001},
archivePrefix = {arXiv},
       eprint = {1306.5395},
 primaryClass = {gr-qc},
       adsurl = {https://ui.adsabs.harvard.edu/abs/2013PhRvD..88h4001T}
}

@ARTICLE{NG15aniso,
       author = {{Agazie}, Gabriella and {Anumarlapudi}, Akash and {Archibald}, Anne M. and {Arzoumanian}, Zaven and {Baker}, Paul T. and {B{\'e}csy}, Bence and {Blecha}, Laura and {Brazier}, Adam and {Brook}, Paul R. and {Burke-Spolaor}, Sarah and {Casey-Clyde}, J. Andrew and {Charisi}, Maria and {Chatterjee}, Shami and {Cohen}, Tyler and {Cordes}, James M. and {Cornish}, Neil J. and {Crawford}, Fronefield and {Cromartie}, H. Thankful and {Crowter}, Kathryn and {DeCesar}, Megan E. and {Demorest}, Paul B. and {Dolch}, Timothy and {Drachler}, Brendan and {Ferrara}, Elizabeth C. and {Fiore}, William and {Fonseca}, Emmanuel and {Freedman}, Gabriel E. and {Gardiner}, Emiko and {Garver-Daniels}, Nate and {Gentile}, Peter A. and {Glaser}, Joseph and {Good}, Deborah C. and {G{\"u}ltekin}, Kayhan and {Hazboun}, Jeffrey S. and {Jennings}, Ross J. and {Johnson}, Aaron D. and {Jones}, Megan L. and {Kaiser}, Andrew R. and {Kaplan}, David L. and {Kelley}, Luke Zoltan and {Kerr}, Matthew and {Key}, Joey S. and {Laal}, Nima and {Lam}, Michael T. and {Lamb}, William G. and {Lazio}, T. Joseph W. and {Lewandowska}, Natalia and {Liu}, Tingting and {Lorimer}, Duncan R. and {Luo}, Jing and {Lynch}, Ryan S. and {Ma}, Chung-Pei and {Madison}, Dustin R. and {McEwen}, Alexander and {McKee}, James W. and {McLaughlin}, Maura A. and {McMann}, Natasha and {Meyers}, Bradley W. and {Mingarelli}, Chiara M.~F. and {Mitridate}, Andrea and {Ng}, Cherry and {Nice}, David J. and {Ocker}, Stella Koch and {Olum}, Ken D. and {Pennucci}, Timothy T. and {Perera}, Benetge B.~P. and {Pol}, Nihan S. and {Radovan}, Henri A. and {Ransom}, Scott M. and {Ray}, Paul S. and {Romano}, Joseph D. and {Sardesai}, Shashwat C. and {Schmiedekamp}, Ann and {Schmiedekamp}, Carl and {Schmitz}, Kai and {Schult}, Levi and {Shapiro-Albert}, Brent J. and {Siemens}, Xavier and {Simon}, Joseph and {Siwek}, Magdalena S. and {Stairs}, Ingrid H. and {Stinebring}, Daniel R. and {Stovall}, Kevin and {Susobhanan}, Abhimanyu and {Swiggum}, Joseph K. and {Taylor}, Stephen R. and {Turner}, Jacob E. and {Unal}, Caner and {Vallisneri}, Michele and {Vigeland}, Sarah J. and {Wahl}, Haley M. and {Witt}, Caitlin A. and {Young}, Olivia},
        title = "{The NANOGrav 15 yr Data Set: Search for Anisotropy in the Gravitational-wave Background}",
      journal = {\apjl},
         year = 2023,
        month = oct,
       volume = {956},
       number = {1},
          eid = {L3},
        pages = {L3},
          doi = {10.3847/2041-8213/acf4fd},
archivePrefix = {arXiv},
       eprint = {2306.16221},
 primaryClass = {astro-ph.HE},
       adsurl = {https://ui.adsabs.harvard.edu/abs/2023ApJ...956L...3A}
}

@ARTICLE{Grunthal_2024,
       author = {{Grunthal}, Kathrin and {Nathan}, Rowina S. and {Thrane}, Eric and {Champion}, David J. and {Miles}, Matthew T. and {Shannon}, Ryan M. and {Kulkarni}, Atharva D. and {Abbate}, Federico and {Buchner}, Sarah and {Cameron}, Andrew D. and {Geyer}, Marisa and {Gitika}, Pratyasha and {Keith}, Michael J. and {Kramer}, Michael and {Lasky}, Paul D. and {Parthasarathy}, Aditya and {Reardon}, Daniel J. and {Singha}, Jaikhomba and {Venkatraman Krishnan}, Vivek},
        title = "{The MeerKAT Pulsar Timing Array: Maps of the gravitational wave sky with the 4.5-yr data release}",
      journal = {\mnras},
         year = 2025,
        month = jan,
       volume = {536},
       number = {2},
        pages = {1501-1517},
          doi = {10.1093/mnras/stae2573},
archivePrefix = {arXiv},
       eprint = {2412.01214},
 primaryClass = {astro-ph.HE},
       adsurl = {https://ui.adsabs.harvard.edu/abs/2025MNRAS.536.1501G}
}

@ARTICLE{depta2024,
       author = {{Depta}, Paul Frederik and {Domcke}, Valerie and {Franciolini}, Gabriele and {Pieroni}, Mauro},
        title = "{Pulsar timing array sensitivity to anisotropies in the gravitational wave background}",
      journal = {\prd},
         year = 2025,
        month = apr,
       volume = {111},
       number = {8},
          eid = {083039},
        pages = {083039},
          doi = {10.1103/PhysRevD.111.083039},
archivePrefix = {arXiv},
       eprint = {2407.14460},
 primaryClass = {astro-ph.CO},
       adsurl = {https://ui.adsabs.harvard.edu/abs/2025PhRvD.111h3039D}
}

@ARTICLE{EPTA_5,
       author = {{EPTA Collaboration} and {InPTA Collaboration} and {Antoniadis}, J. and {Arumugam}, P. and {Arumugam}, S. and {Babak}, S. and {Bagchi}, M. and {Bak Nielsen}, A.-S. and {Bassa}, C.~G. and {Bathula}, A. and {Berthereau}, A. and {Bonetti}, M. and {Bortolas}, E. and {Brook}, P.~R. and {Burgay}, M. and {Caballero}, R.~N. and {Chalumeau}, A. and {Champion}, D.~J. and {Chanlaridis}, S. and {Chen}, S. and {Cognard}, I. and {Dandapat}, S. and {Deb}, D. and {Desai}, S. and {Desvignes}, G. and {Dhanda-Batra}, N. and {Dwivedi}, C. and {Falxa}, M. and {Ferranti}, I. and {Ferdman}, R.~D. and {Franchini}, A. and {Gair}, J.~R. and {Goncharov}, B. and {Gopakumar}, A. and {Graikou}, E. and {Grie{\ss}meier}, J.-M. and {Guillemot}, L. and {Guo}, Y.~J. and {Gupta}, Y. and {Hisano}, S. and {Hu}, H. and {Iraci}, F. and {Izquierdo-Villalba}, D. and {Jang}, J. and {Jawor}, J. and {Janssen}, G.~H. and {Jessner}, A. and {Joshi}, B.~C. and {Kareem}, F. and {Karuppusamy}, R. and {Keane}, E.~F. and {Keith}, M.~J. and {Kharbanda}, D. and {Kikunaga}, T. and {Kolhe}, N. and {Kramer}, M. and {Krishnakumar}, M.~A. and {Lackeos}, K. and {Lee}, K.~J. and {Liu}, K. and {Liu}, Y. and {Lyne}, A.~G. and {McKee}, J.~W. and {Maan}, Y. and {Main}, R.~A. and {Manzini}, S. and {Mickaliger}, M.~B. and {Ni{\c{t}}u}, I.~C. and {Nobleson}, K. and {Paladi}, A.~K. and {Parthasarathy}, A. and {Perera}, B.~B.~P. and {Perrodin}, D. and {Petiteau}, A. and {Porayko}, N.~K. and {Possenti}, A. and {Prabu}, T. and {Quelquejay Leclere}, H. and {Rana}, P. and {Samajdar}, A. and {Sanidas}, S.~A. and {Sesana}, A. and {Shaifullah}, G. and {Singha}, J. and {Speri}, L. and {Spiewak}, R. and {Srivastava}, A. and {Stappers}, B.~W. and {Surnis}, M. and {Susarla}, S.~C. and {Susobhanan}, A. and {Takahashi}, K. and {Tarafdar}, P. and {Theureau}, G. and {Tiburzi}, C. and {van der Wateren}, E. and {Vecchio}, A. and {Venkatraman Krishnan}, V. and {Verbiest}, J.~P.~W. and {Wang}, J. and {Wang}, L. and {Wu}, Z.},
        title = "{The second data release from the European Pulsar Timing Array. V. Search for continuous gravitational wave signals}",
      journal = {\aap},
         year = 2024,
        month = oct,
       volume = {690},
          eid = {A118},
        pages = {A118},
          doi = {10.1051/0004-6361/202348568},
archivePrefix = {arXiv},
       eprint = {2306.16226},
 primaryClass = {astro-ph.HE},
       adsurl = {https://ui.adsabs.harvard.edu/abs/2024A&A...690A.118E}
}

@ARTICLE{ng15yrCGW,
       author = {{Agazie}, Gabriella and {Anumarlapudi}, Akash and {Archibald}, Anne M. and {Arzoumanian}, Zaven and {Baker}, Paul T. and {B{\'e}csy}, Bence and {Blecha}, Laura and {Brazier}, Adam and {Brook}, Paul R. and {Burke-Spolaor}, Sarah and {Case}, Robin and {Casey-Clyde}, J. Andrew and {Charisi}, Maria and {Chatterjee}, Shami and {Cohen}, Tyler and {Cordes}, James M. and {Cornish}, Neil J. and {Crawford}, Fronefield and {Cromartie}, H. Thankful and {Crowter}, Kathryn and {Decesar}, Megan E. and {Demorest}, Paul B. and {Digman}, Matthew C. and {Dolch}, Timothy and {Drachler}, Brendan and {Ferrara}, Elizabeth C. and {Fiore}, William and {Fonseca}, Emmanuel and {Freedman}, Gabriel E. and {Garver-Daniels}, Nate and {Gentile}, Peter A. and {Glaser}, Joseph and {Good}, Deborah C. and {G{\"u}ltekin}, Kayhan and {Hazboun}, Jeffrey S. and {Hourihane}, Sophie and {Jennings}, Ross J. and {Johnson}, Aaron D. and {Jones}, Megan L. and {Kaiser}, Andrew R. and {Kaplan}, David L. and {Kelley}, Luke Zoltan and {Kerr}, Matthew and {Key}, Joey S. and {Laal}, Nima and {Lam}, Michael T. and {Lamb}, William G. and {Lazio}, T. Joseph W. and {Lewandowska}, Natalia and {Liu}, Tingting and {Lorimer}, Duncan R. and {Luo}, Jing and {Lynch}, Ryan S. and {Ma}, Chung-Pei and {Madison}, Dustin R. and {McEwen}, Alexander and {McKee}, James W. and {McLaughlin}, Maura A. and {McMann}, Natasha and {Meyers}, Bradley W. and {Meyers}, Patrick M. and {Mingarelli}, Chiara M.~F. and {Mitridate}, Andrea and {Ng}, Cherry and {Nice}, David J. and {Ocker}, Stella Koch and {Olum}, Ken D. and {Pennucci}, Timothy T. and {Perera}, Benetge B.~P. and {Petrov}, Polina and {Pol}, Nihan S. and {Radovan}, Henri A. and {Ransom}, Scott M. and {Ray}, Paul S. and {Romano}, Joseph D. and {Sardesai}, Shashwat C. and {Schmiedekamp}, Ann and {Schmiedekamp}, Carl and {Schmitz}, Kai and {Shapiro-Albert}, Brent J. and {Siemens}, Xavier and {Simon}, Joseph and {Siwek}, Magdalena S. and {Stairs}, Ingrid H. and {Stinebring}, Daniel R. and {Stovall}, Kevin and {Susobhanan}, Abhimanyu and {Swiggum}, Joseph K. and {Taylor}, Jacob and {Taylor}, Stephen R. and {Turner}, Jacob E. and {Unal}, Caner and {Vallisneri}, Michele and {van Haasteren}, Rutger and {Vigeland}, Sarah J. and {Wahl}, Haley M. and {Witt}, Caitlin A. and {Young}, Olivia and {Nanograv Collaboration}},
        title = "{The NANOGrav 15 yr Data Set: Bayesian Limits on Gravitational Waves from Individual Supermassive Black Hole Binaries}",
      journal = {\apjl},
         year = 2023,
        month = jul,
       volume = {951},
       number = {2},
          eid = {L50},
        pages = {L50},
          doi = {10.3847/2041-8213/ace18a},
archivePrefix = {arXiv},
       eprint = {2306.16222},
 primaryClass = {astro-ph.HE},
       adsurl = {https://ui.adsabs.harvard.edu/abs/2023ApJ...951L..50A}
}

@ARTICLE{bence_exp,
       author = {{B{\'e}csy}, Bence and {Cornish}, Neil J. and {Kelley}, Luke Zoltan},
        title = "{Exploring Realistic Nanohertz Gravitational-wave Backgrounds}",
      journal = {\apj},
         year = 2022,
        month = dec,
       volume = {941},
       number = {2},
          eid = {119},
        pages = {119},
          doi = {10.3847/1538-4357/aca1b2},
archivePrefix = {arXiv},
       eprint = {2207.01607},
 primaryClass = {astro-ph.HE},
       adsurl = {https://ui.adsabs.harvard.edu/abs/2022ApJ...941..119B}
}

@ARTICLE{2026PhRvD.113j3031G,
       author = {{Gersbach}, Kyle A. and {Taylor}, Stephen R. and {B{\'e}csy}, Bence and {Lemke}, Anna-Malin and {Mitridate}, Andrea and {Pol}, Nihan},
        title = "{Mapping the gravitational-wave background across the spectrum with a next-generation anisotropic per-frequency optimal statistic}",
      journal = {\prd},
         year = 2026,
        month = may,
       volume = {113},
       number = {10},
          eid = {103031},
        pages = {103031},
          doi = {10.1103/d55k-75p9},
archivePrefix = {arXiv},
       eprint = {2509.07090},
 primaryClass = {astro-ph.IM},
       adsurl = {https://ui.adsabs.harvard.edu/abs/2026PhRvD.113j3031G}
}

@software{2025zndo..17711453V,
       author = {{Vallisneri}, Michele and {Meyers}, Patrick M. and {Wright}, David and {Johnson}, Aaron D. and {Baier}, Jeremy G. and {van Haasteren}, Rutger},
        title = "{nanograv/discovery: 0.5.1}",
         year = 2025,
        month = nov,
          eid = {10.5281/zenodo.17711453},
          doi = {10.5281/zenodo.17711453},
      version = {0.5.1},
    publisher = {Zenodo},
       adsurl = {https://ui.adsabs.harvard.edu/abs/2025zndo..17711453V},
 howpublished = {\url{https://doi.org/10.5281/zenodo.17711453}}
}

@ARTICLE{2025PhRvL.135g1401V,
       author = {{Vallisneri}, Michele and {Crisostomi}, Marco and {Johnson}, Aaron D. and {Meyers}, Patrick M.},
        title = "{Rapid Parameter Estimation for Pulsar-Timing-Array Datasets with Variational Inference and Normalizing Flows}",
      journal = {\prl},
         year = 2025,
        month = aug,
       volume = {135},
       number = {7},
          eid = {071401},
        pages = {071401},
          doi = {10.1103/p3f7-rbmv},
archivePrefix = {arXiv},
       eprint = {2405.08857},
 primaryClass = {gr-qc},
       adsurl = {https://ui.adsabs.harvard.edu/abs/2025PhRvL.135g1401V}
}

@ARTICLE{2025arXiv250613866C,
       author = {{Crisostomi}, Marco and {van Haasteren}, Rutger and {Meyers}, Patrick M. and {Vallisneri}, Michele},
        title = "{Beyond diagonal approximations: improved covariance modeling for pulsar timing array data analysis}",
      journal = {arXiv e-prints},
         year = 2025,
        month = jun,
          eid = {arXiv:2506.13866},
        pages = {arXiv:2506.13866},
          doi = {10.48550/arXiv.2506.13866},
archivePrefix = {arXiv},
       eprint = {2506.13866},
 primaryClass = {astro-ph.IM},
       adsurl = {https://ui.adsabs.harvard.edu/abs/2025arXiv250613866C}
}

@ARTICLE{2024ApJ...976..212S,
       author = {{Sardesai}, Shashwat C. and {Simon}, Joseph and {Vigeland}, Sarah J.},
        title = "{Characterizing the Nanohertz Gravitational Wave Background Using a t-process Power Spectral Density}",
      journal = {\apj},
         year = 2024,
        month = dec,
       volume = {976},
       number = {2},
          eid = {212},
        pages = {212},
          doi = {10.3847/1538-4357/ad8a60},
archivePrefix = {arXiv},
       eprint = {2408.10139},
 primaryClass = {astro-ph.HE},
       adsurl = {https://ui.adsabs.harvard.edu/abs/2024ApJ...976..212S}
}

@ARTICLE{2023ApJ...951L...8A,
       author = {{Agazie}, Gabriella and {Anumarlapudi}, Akash and {Archibald}, Anne M. and {Arzoumanian}, Zaven and {Baker}, Paul T. and {B{\'e}csy}, Bence and {Blecha}, Laura and {Brazier}, Adam and {Brook}, Paul R. and {Burke-Spolaor}, Sarah and {Burnette}, Rand and {Case}, Robin and {Charisi}, Maria and {Chatterjee}, Shami and {Chatziioannou}, Katerina and {Cheeseboro}, Belinda D. and {Chen}, Siyuan and {Cohen}, Tyler and {Cordes}, James M. and {Cornish}, Neil J. and {Crawford}, Fronefield and {Cromartie}, H. Thankful and {Crowter}, Kathryn and {Cutler}, Curt J. and {Decesar}, Megan E. and {Degan}, Dallas and {Demorest}, Paul B. and {Deng}, Heling and {Dolch}, Timothy and {Drachler}, Brendan and {Ellis}, Justin A. and {Ferrara}, Elizabeth C. and {Fiore}, William and {Fonseca}, Emmanuel and {Freedman}, Gabriel E. and {Garver-Daniels}, Nate and {Gentile}, Peter A. and {Gersbach}, Kyle A. and {Glaser}, Joseph and {Good}, Deborah C. and {G{\"u}ltekin}, Kayhan and {Hazboun}, Jeffrey S. and {Hourihane}, Sophie and {Islo}, Kristina and {Jennings}, Ross J. and {Johnson}, Aaron D. and {Jones}, Megan L. and {Kaiser}, Andrew R. and {Kaplan}, David L. and {Kelley}, Luke Zoltan and {Kerr}, Matthew and {Key}, Joey S. and {Klein}, Tonia C. and {Laal}, Nima and {Lam}, Michael T. and {Lamb}, William G. and {Lazio}, T. Joseph W. and {Lewandowska}, Natalia and {Littenberg}, Tyson B. and {Liu}, Tingting and {Lommen}, Andrea and {Lorimer}, Duncan R. and {Luo}, Jing and {Lynch}, Ryan S. and {Ma}, Chung-Pei and {Madison}, Dustin R. and {Mattson}, Margaret A. and {McEwen}, Alexander and {McKee}, James W. and {McLaughlin}, Maura A. and {McMann}, Natasha and {Meyers}, Bradley W. and {Meyers}, Patrick M. and {Mingarelli}, Chiara M.~F. and {Mitridate}, Andrea and {Natarajan}, Priyamvada and {Ng}, Cherry and {Nice}, David J. and {Ocker}, Stella Koch and {Olum}, Ken D. and {Pennucci}, Timothy T. and {Perera}, Benetge B.~P. and {Petrov}, Polina and {Pol}, Nihan S. and {Radovan}, Henri A. and {Ransom}, Scott M. and {Ray}, Paul S. and {Romano}, Joseph D. and {Sardesai}, Shashwat C. and {Schmiedekamp}, Ann and {Schmiedekamp}, Carl and {Schmitz}, Kai and {Schult}, Levi and {Shapiro-Albert}, Brent J. and {Siemens}, Xavier and {Simon}, Joseph and {Siwek}, Magdalena S. and {Stairs}, Ingrid H. and {Stinebring}, Daniel R. and {Stovall}, Kevin and {Sun}, Jerry P. and {Susobhanan}, Abhimanyu and {Swiggum}, Joseph K. and {Taylor}, Jacob and {Taylor}, Stephen R. and {Turner}, Jacob E. and {Unal}, Caner and {Vallisneri}, Michele and {van Haasteren}, Rutger and {Vigeland}, Sarah J. and {Wahl}, Haley M. and {Wang}, Qiaohong and {Witt}, Caitlin A. and {Young}, Olivia and {Nanograv Collaboration}},
        title = "{The NANOGrav 15 yr Data Set: Evidence for a Gravitational-wave Background}",
      journal = {\apjl},
         year = 2023,
        month = jul,
       volume = {951},
       number = {1},
          eid = {L8},
        pages = {L8},
          doi = {10.3847/2041-8213/acdac6},
archivePrefix = {arXiv},
       eprint = {2306.16213},
 primaryClass = {astro-ph.HE},
       adsurl = {https://ui.adsabs.harvard.edu/abs/2023ApJ...951L...8A}
}

@ARTICLE{2020ApJ...900..102A,
       author = {{Arzoumanian}, Zaven and {Baker}, Paul T. and {Brazier}, Adam and {Brook}, Paul R. and {Burke-Spolaor}, Sarah and {B{\'e}csy}, Bence and {Charisi}, Maria and {Chatterjee}, Shami and {Cordes}, James M. and {Cornish}, Neil J. and {Crawford}, Fronefield and {Cromartie}, H. Thankful and {Crowter}, Kathryn and {Decesar}, Megan E. and {Demorest}, Paul B. and {Dolch}, Timothy and {Elliott}, Rodney D. and {Ellis}, Justin A. and {Ferdman}, Robert D. and {Ferrara}, Elizabeth C. and {Fonseca}, Emmanuel and {Garver-Daniels}, Nathan and {Gentile}, Peter A. and {Good}, Deborah C. and {Hazboun}, Jeffrey S. and {Islo}, Kristina and {Jennings}, Ross J. and {Jones}, Megan L. and {Kaiser}, Andrew R. and {Kaplan}, David L. and {Kelley}, Luke Zoltan and {Key}, Joey Shapiro and {Lam}, Michael T. and {Lazio}, T. Joseph W. and {Levin}, Lina and {Luo}, Jing and {Lynch}, Ryan S. and {Madison}, Dustin R. and {McLaughlin}, Maura A. and {Mingarelli}, Chiara M.~F. and {Ng}, Cherry and {Nice}, David J. and {Pennucci}, Timothy T. and {Pol}, Nihan S. and {Ransom}, Scott M. and {Ray}, Paul S. and {Shapiro-Albert}, Brent J. and {Siemens}, Xavier and {Simon}, Joseph and {Spiewak}, Ren{\'e}e and {Stairs}, Ingrid H. and {Stinebring}, Daniel R. and {Stovall}, Kevin and {Swiggum}, Joseph K. and {Taylor}, Stephen R. and {Vallisneri}, Michele and {Vigeland}, Sarah J. and {Witt}, Caitlin A. and {Zhu}, Weiwei and {NANOGrav Collaboration}},
        title = "{Multimessenger Gravitational-wave Searches with Pulsar Timing Arrays: Application to 3C 66B Using the NANOGrav 11-year Data Set}",
      journal = {\apj},
         year = 2020,
        month = sep,
       volume = {900},
       number = {2},
          eid = {102},
        pages = {102},
          doi = {10.3847/1538-4357/ababa1},
archivePrefix = {arXiv},
       eprint = {2005.07123},
 primaryClass = {astro-ph.GA},
       adsurl = {https://ui.adsabs.harvard.edu/abs/2020ApJ...900..102A}
}

@ARTICLE{2019ApJ...880..116A,
       author = {{Aggarwal}, K. and {Arzoumanian}, Z. and {Baker}, P.~T. and {Brazier}, A. and {Brinson}, M.~R. and {Brook}, P.~R. and {Burke-Spolaor}, S. and {Chatterjee}, S. and {Cordes}, J.~M. and {Cornish}, N.~J. and {Crawford}, F. and {Crowter}, K. and {Cromartie}, H.~T. and {DeCesar}, M. and {Demorest}, P.~B. and {Dolch}, T. and {Ellis}, J.~A. and {Ferdman}, R.~D. and {Ferrara}, E. and {Fonseca}, E. and {Garver-Daniels}, N. and {Gentile}, P. and {Hazboun}, J.~S. and {Holgado}, A.~M. and {Huerta}, E.~A. and {Islo}, K. and {Jennings}, R. and {Jones}, G. and {Jones}, M.~L. and {Kaiser}, A.~R. and {Kaplan}, D.~L. and {Kelley}, L.~Z. and {Key}, J.~S. and {Lam}, M.~T. and {Lazio}, T.~J.~W. and {Levin}, L. and {Lorimer}, D.~R. and {Luo}, J. and {Lynch}, R.~S. and {Madison}, D.~R. and {McLaughlin}, M.~A. and {McWilliams}, S.~T. and {Mingarelli}, C.~M.~F. and {Ng}, C. and {Nice}, D.~J. and {Pennucci}, T.~T. and {Pol}, N.~S. and {Ransom}, S.~M. and {Ray}, P.~S. and {Siemens}, X. and {Simon}, J. and {Spiewak}, R. and {Stairs}, I.~H. and {Stinebring}, D.~R. and {Stovall}, K. and {Swiggum}, J. and {Taylor}, S.~R. and {Turner}, J.~E. and {Vallisneri}, M. and {van Haasteren}, R. and {Vigeland}, S.~J. and {Witt}, C.~A. and {Zhu}, W.~W. and {NANOGrav Collaboration}},
        title = "{The NANOGrav 11 yr Data Set: Limits on Gravitational Waves from Individual Supermassive Black Hole Binaries}",
      journal = {\apj},
         year = 2019,
        month = aug,
       volume = {880},
       number = {2},
          eid = {116},
        pages = {116},
          doi = {10.3847/1538-4357/ab2236},
archivePrefix = {arXiv},
       eprint = {1812.11585},
 primaryClass = {astro-ph.GA},
       adsurl = {https://ui.adsabs.harvard.edu/abs/2019ApJ...880..116A}
}

@ARTICLE{2018PhRvD..98d4003V,
       author = {{Vigeland}, Sarah J. and {Islo}, Kristina and {Taylor}, Stephen R. and {Ellis}, Justin A.},
        title = "{Noise-marginalized optimal statistic: A robust hybrid frequentist-Bayesian statistic for the stochastic gravitational-wave background in pulsar timing arrays}",
      journal = {\prd},
         year = 2018,
        month = aug,
       volume = {98},
       number = {4},
          eid = {044003},
        pages = {044003},
          doi = {10.1103/PhysRevD.98.044003},
archivePrefix = {arXiv},
       eprint = {1805.12188},
 primaryClass = {astro-ph.IM},
       adsurl = {https://ui.adsabs.harvard.edu/abs/2018PhRvD..98d4003V}
}

@ARTICLE{1976JPhA....9.1387K,
       author = {{Kibble}, T.~W.~B.},
        title = "{Topology of cosmic domains and strings}",
      journal = {Journal of Physics A Mathematical General},
         year = 1976,
        month = aug,
       volume = {9},
       number = {8},
        pages = {1387-1398},
          doi = {10.1088/0305-4470/9/8/029},
       adsurl = {https://ui.adsabs.harvard.edu/abs/1976JPhA....9.1387K}
}

@ARTICLE{2024MNRAS.532..295F,
       author = {{Fastidio}, F. and {Gualandris}, A. and {Sesana}, A. and {Bortolas}, E. and {Dehnen}, W.},
        title = "{Eccentricity evolution of PTA sources from cosmological initial conditions}",
      journal = {\mnras},
         year = 2024,
        month = jul,
       volume = {532},
       number = {1},
        pages = {295-304},
          doi = {10.1093/mnras/stae1411},
archivePrefix = {arXiv},
       eprint = {2406.02710},
 primaryClass = {astro-ph.GA},
       adsurl = {https://ui.adsabs.harvard.edu/abs/2024MNRAS.532..295F}
}

@ARTICLE{vHV14,
       author = {{van Haasteren}, Rutger and {Vallisneri}, Michele},
        title = "{New advances in the Gaussian-process approach to pulsar-timing data analysis}",
      journal = {\prd},
         year = 2014,
        month = nov,
       volume = {90},
       number = {10},
          eid = {104012},
        pages = {104012},
          doi = {10.1103/PhysRevD.90.104012},
archivePrefix = {arXiv},
       eprint = {1407.1838},
 primaryClass = {gr-qc},
       adsurl = {https://ui.adsabs.harvard.edu/abs/2014PhRvD..90j4012V}
}

\end{document}